\documentclass[aps,twocolumn,showpacs,pre,UTF8]{revtex4}
\usepackage{amsmath,epsfig,amssymb,subfigure,bm,dsfont}
\usepackage{amsmath}
\usepackage{txfonts}
\usepackage[titletoc]{appendix}
\usepackage{amssymb}
\usepackage{array}
\usepackage{mathrsfs}
\usepackage{subfigure,overpic}
\usepackage{dcolumn}
\usepackage{graphicx}
\usepackage{epstopdf}
\usepackage{float}
\usepackage{flushend}
\usepackage{balance}
\usepackage{lipsum}
\usepackage[colorlinks,
            linkcolor=blue,
            anchorcolor=blue,
            citecolor=blue]{hyperref}

\begin{document}

\title{RKKY signals characterizing the topological phase transitions in Floquet Dirac semimetals}
\author{Hou-Jian Duan}
\email{dhjphd@163.com}
\author{Shi-Ming Cai}
\author{Xing Wei}
\author{Yong-Chi Chen}
\author{Yong-Jia Wu}
\author{Ming-Xun Deng}
\author{Ruiqiang Wang}
\email{wangruiqiang@m.scnu.edu.cn}
\author{Mou Yang}
\affiliation{Guangdong Basic Research Center of Excellence for Structure and Fundamental Interactions of Matter, Guangdong Provincial Key Laboratory of Quantum Engineering and Quantum Materials, School of Physics, South China Normal University, Guangzhou 510006, China}
\affiliation{Guangdong-Hong Kong Joint Laboratory of Quantum Matter, Frontier Research Institute for Physics, South China Normal University, Guangzhou 510006, China}

\begin{abstract}
Recently, the Floquet ${\rm Na_3Bi}$-type material has been proposed as an ideal platform for realizing various phases, i.e., the spin-degenerate Dirac semimetal (DSM) can be turned into the Weyl semimetal (WSM), and even to the Weyl half-metal (WHM)\cite{xiaoshi}. Instead of the conventional electrical methods, we use the RKKY interaction to characterize the topological phase transitions in this paper. It is found that detecting the Ising term $J_I$ is feasible for distinguishing the phase transition of DSM/WSM, since the emergence of $J_I$ is induced by the broken spin degeneracy. For the case with impurities deposited on $z$ axis (the line connecting the Weyl points), the Heisenberg term $J_H$ coexists with $J_I$ in the WSM, while $J_H$ is filtered out and only $J_I$ survives in the WHM. This magnetic filtering effect is a reflection of the fully spin-polarized property (one spin band is in the WSM phase while the other is gapped) of the WHM, and it can act a signal to capture the phase transition of WSM/WHM. This signal can not be disturbed unless the direction of the impurities greatly deviates from $z$ axis. Interestingly, as the impurities are moved into the $x$-$y$ plane, there arises another signal (a dip structure for $J_H$ at the phase boundary), which can also identify the phase transition of WSM/WHM. Furthermore, we have verified that all magnetic signals are robust to the term that breaks the electron-hole symmetry. Besides characterizing the phase transitions, our results also suggest that the Floquet DSMs are power platforms for controlling the magnetic interaction.
\end{abstract}

\maketitle

\section{introduction}
The study of the topological states has become a hot spot in condensed matter and it recently excites a great interest in realizing various topological phases, including topological insulators and topological semimetals. One powerful method to generate topological states is to apply electromagnetic radiation\cite{F0,F1,F2,F3,F4,F5,F6,F7,F8}, which can rearrange the band structure and change material properties by photon dressing. For example, topological Weyl semimetals (WSMs) can be obtained by applying a beam of circularly polarized light (CPL) in nodal-line semimetals\cite{F6} or Dirac semimetals (DSMs)\cite{F7,F8}. Usually, these so-called Floquet topological states can be controlled by the light intensity (or frequency), and the optical tunability offers the related materials a great potential for applications in spintronics. Remarkably, the Floquet topological states have been experimentally realized in artificial photonic lattices\cite{E1}, as well as in the solid\cite{E2}.
\par
Recently, the ${\rm Na_3Bi}$-type DSM has attracted us due to the various topological phases induced by the off-resonant CPL\cite{xiaoshi}. As stated in Ref. \cite{xiaoshi}, the original DSM is changed to be a WSM once the light is turned on. More interestingly, besides the WSM, the Weyl half-metal (WHM) can be obtained if the light intensity exceeds a critical value. Compared to the WSM, the WHM possesses greater potential for the development of spintronic devices since it acts as a perfect spin filter in the Dirac-Weyl semimetal junction. This transport property is resulting from the fully spin-polarized property of the WHM, i.e., one spin band is in the WSM phase while the other is in the insulator phase. The discovery of the various phases in Floquet DSMs raises an interesting topic: how to detect these topological phases? To solve this problem, a conventional method is to measure the spin-resolved quantum Hall conductivity or probe the surface states directly. However, the accuracy of these methods is highly dependent on the purity of the materials, since impurities or defects are unavoidable in real materials. Moreover, the surface states are susceptible to the disturbance from bulk states in topological semimetals. Thus, new methods for probing the phase transitions are necessary.
\par
 The RKKY interaction between magnetic impurities offers the possibility for detecting the phase transitions, since it is sensitive to the deformation of the band structure of the materials. Typically, magnetic signals can be extracted from the amplitude, the oscillation and the decaying laws of the RKKY interaction for characterizing the properties of the materials. For examples, the amplitude of the RKKY interaction contributed by the edge states is about twenty times greater than the bulk contribution in the silicene nanoribbon\cite{surface1}, a sharp peak or dip of the RKKY interaction at the phase boundary can be applied to identify the phase transition from the type-I WSM to the type-II WSM\cite{Weyl3}, the significant difference in the amplitude of the RKKY interaction can act as a signal to distinguish between the topological Fermi surface and the trivial Fermi surface in nodal-line semimetal\cite{NLSM}, the anisotropic decaying laws of the RKKY interaction can be used as the evidence that the semi-DSMs (S-DSMs) are distinct from other isotropic systems\cite{semiDirac}, the splitting of the Weyl points in WSMs can be captured by the oscillation of the RKKY interaction\cite{rkky1,rkky2,rkky3}. Besides the static systems, various signals are also extracted from the RKKY interaction for characterizing the properties of the Floquet band structure in irradiated systems\cite{floquetRKKY3,floquetRKKY1,floquetRKKY2}.
\par
In this paper, by doping magnetic impurities in the bulk of the Floquet ${\rm Na_3Bi}$-type DSM, it is expected that signals can be extracted from the RKKY interaction to characterize the various phase transitions. Considering the off-resonant condition of the CPL, the RKKY interaction can be calculated with the aid of the static (time-independent) Green's function\cite{floquetRKKY1,floquetRKKY2}. It is found that the RKKY components have completely different responses to the light parameters in different phases. By checking the Ising term, the phase transition of DSM/WSM can be identified. Depending on the impurity configuration, different signals in the Heisenberg term can be used to ascertain the phase boundary between the WSM and the WHM. Furthermore, the fully spin-polarized property of the WHM can also be reflected on the RKKY interaction. In addition, we have discussed the effect of the broken electron-hole symmetry on the magnetic signals. From these discussions, we have proved that the RKKY interaction can be used as an effective method for probing the phase transitions in ${\rm Na_3Bi}$-type DSMs. Also, we have revealed that the Floquet DSMs are great platforms for controlling the magnetic interaction.
\par
Our paper is organized as follows. In Sec. II, the low-energy model of the Floquet ${\rm Na_3Bi}$-type DSM is introduced, and various phase transitions are exhibited. In addition, the method for calculating the RKKY interaction is raised. In Sec. III, the RKKY interaction in different phases are discussed with impurities placed in different directions. In Sec. IV, the term  breaking the electron-hole symmetry is added and its effects on the magnetic signals are discussed. Finally, a summary is drawn in Sec. V.

\section{Model and Method}
\begin{figure}[!htb]
\centering \includegraphics[width=0.25\textwidth]{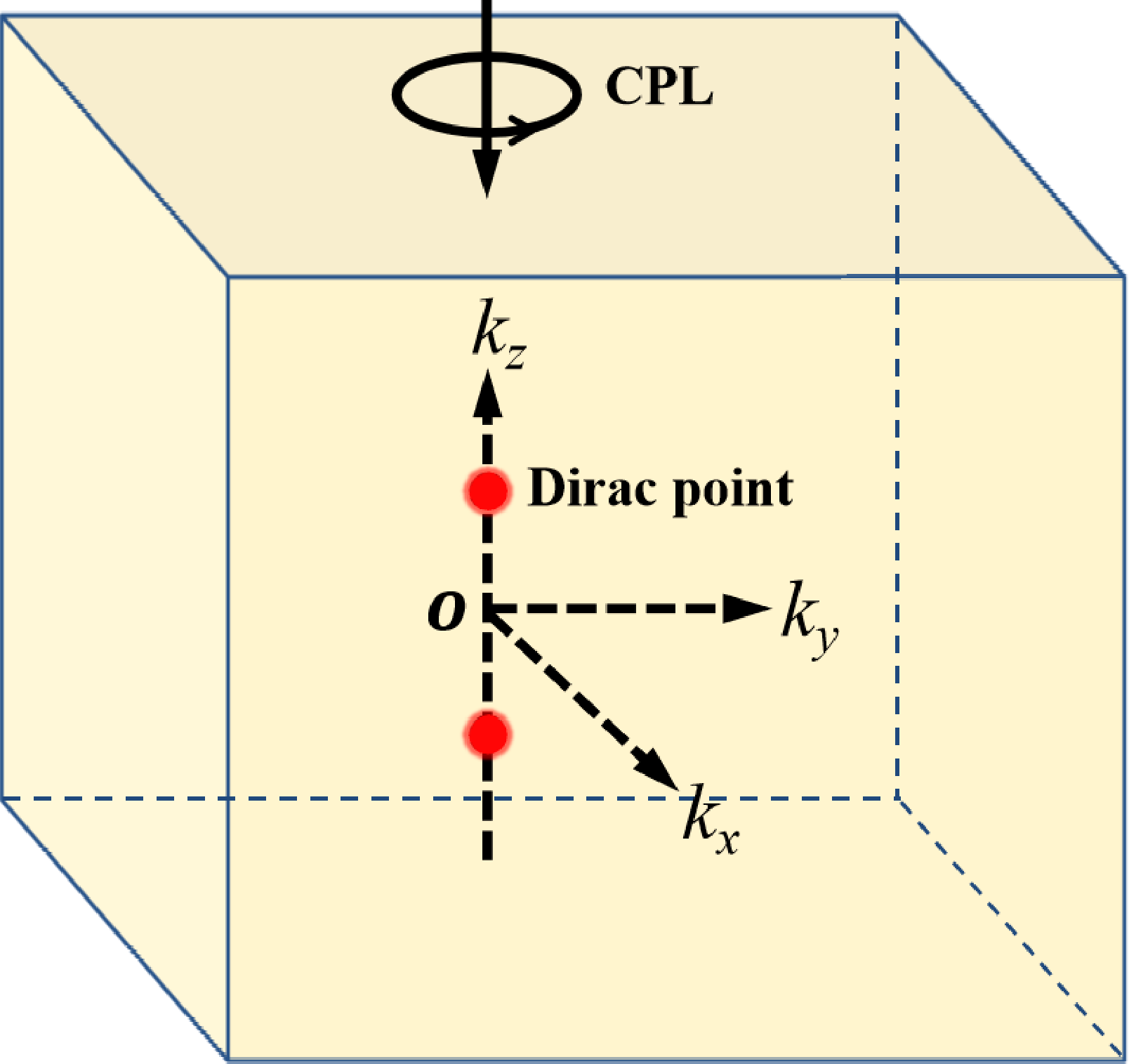}
\caption{(Color online) Schematic of ${\rm Na_3Bi}$-type DSMs with two Dirac points located on $k_z$ axis with positions (0,0,$\pm k_0$) ($k_0=\sqrt{M_0/M_1}$), each of which contains two spin-resolved Weyl points with opposite chiralities. Along $z$ axis, a beam of off-resonant CPL is assumed to be irradiated.}
\end{figure}
We start with a Floquet DSM model introduced in Refs. \cite{ZWang1,ZWang2}, where ${\rm Na_3Bi}$ and ${\rm  Cd_3As_2}$  act as the prototypes of the DSMs. The corresponding low-energy Hamiltonian in orbital and spin basis of $\left(|S,\frac{1}{2}\rangle,|P,\frac{3}{2}\rangle,|S,-\frac{1}{2}\rangle,|P,-\frac{3}{2}\rangle\right)$ can be written as
  \begin{eqnarray}\label{m1}
H\left(\bf{k}\right)=\left(\begin{array}{cc}
\epsilon _{0}\left(\bf{k}\right)+\mathbf{h_{+}}\left(\bf{k}\right)\cdot \mathbf{\tau}&\bf{0}\\
\bf{0}&\epsilon _{0}\left(\bf{k}\right)+\mathbf{h_{-}}\left(\bf{k}\right)\cdot\mathbf{\tau}
\end{array}\right)
\end{eqnarray}
with
\begin{eqnarray}\label{m2}
\begin{split}
\epsilon _{0}\left(\bf{k}\right)=&C'_{0}+C_{1}k_{z}^{2}+C_{2}k_\parallel^2 ,\\
\mathbf{h}_s\left(\bf{k}\right)=& \left(sv_s k_{x},\;-v_{s}k_{y},\;M^\prime_{0}-M_{1}k_{z}^{2}-M_{2}k_\parallel^2-s\lambda\right),
\end{split}
\end{eqnarray}
where $k_\parallel^2=k_x^2+k_y^2$, $C_0^\prime=C_{0}+C_{2}k_{A}^{2}$, $M^\prime_0=M_{0}-M_{2}k_{A}^{2}$, $\lambda=v_0^2k_A^2/(\hbar\Omega)$, and $v_s=v_0-sv_A$ with $v_A=2v_0M_2k_A^2/(\hbar\Omega)$. Here, $k_A$ and $\Omega$ refer to the light intensity and frequency respectively, $\mathbf{\tau}=(\tau_x,\tau_y,\tau_z)$ is the vector of pauli matrix in orbital space, and the subscript $s=+$ ($-$) for spin up (down). Noting that all terms in Eq. (\ref{m2}) related to $k_A$ are induced by applying a beam of off-resonant light to the DSMs (a detailed derivation is given in the Appendix I). The diagonal term $\epsilon _{0}\left(\bf{k}\right)$ in $H\left(\bf{k}\right)$ of Eq. (\ref{m1}) breaks the electron-hole symmetry and $M^\prime_{0}$ in $h_s\left(\bf{k}\right)$ acts as the Dirac mass.
\par
\begin{figure}[!htb]
\centering \includegraphics[width=0.42\textwidth]{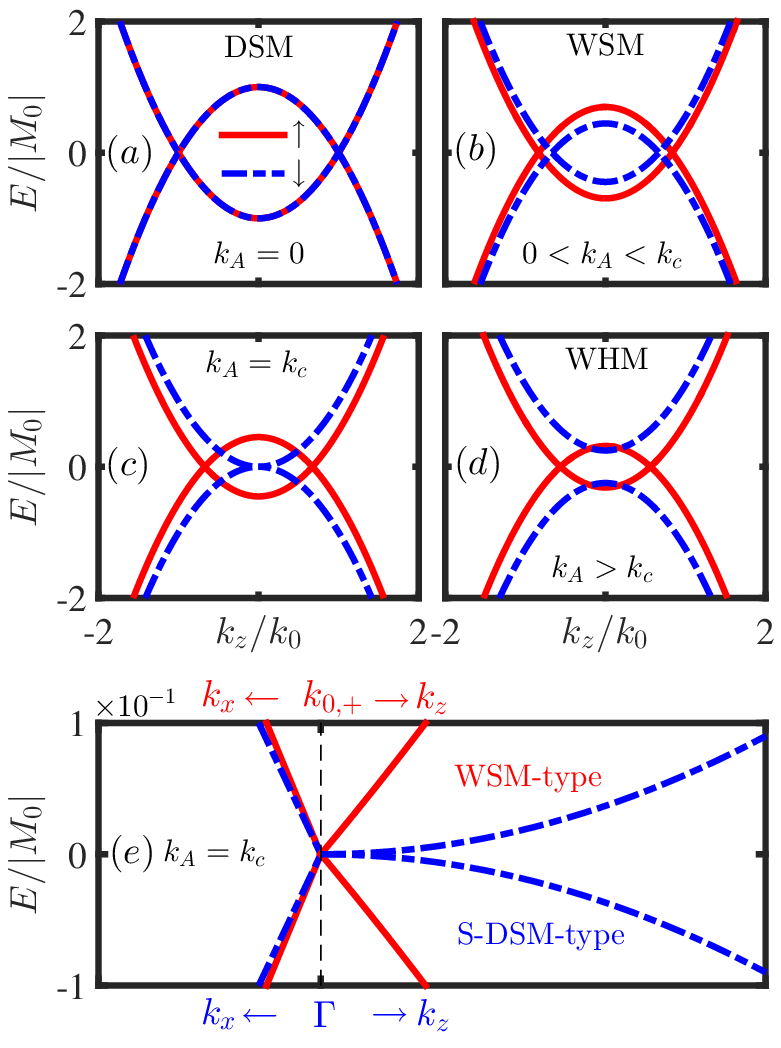}
\caption{(Color online)  Evolution of the $k_z$-axis dispersion with different values of $k_A$, which change the material from (a) DSM to (b) WSM,  and then to (d) WHM. The solid (dashed) lines denote the spin-up (spin-down) bands. (c) The $k_z$-axis dispersion for the phase boundary ($k_A=k_c$) between the WSM and the WHM. The related low-energy dispersion is shown in (e), where the spin-up band around the Weyl point $k_{0,+}$ is linear in all directions while the spin-down band exhibits a semi-Dirac shape around the $\Gamma$ point (i.e., linear in $k_x$ axis but disperses quadratically in $k_z$ axis). Here, $k_{0,+}=\sqrt{(M_0-M_2k_A^2-\lambda)/M_1}$, $k_c=\sqrt{M_0/(M_2- v_0^2/\hbar\Omega)}$ with $\hbar\Omega=2$ ${\rm eV}$ and $\epsilon_0$ is temporarily dropped (i.e., $\epsilon_0=0$).  Parameters $M_0=-0.08686$ ${\rm eV}$, $M_1=-10.6424$ ${\rm eV}$ ${\rm \AA^2}$, $M_2=-10.3610$ ${\rm eV}$ ${\rm \AA^2}$, $v_0=2.4598$ ${\rm eV}$ ${\rm \AA}$ are extracted from ${\rm Na_3Bi}$ \cite{ZWang1} material.}
\end{figure}
By diagonalizing the Hamiltonian of Eq. (\ref{m1}), the energy dispersion can be solved as
\begin{eqnarray}\label{m3}
\begin{split}
E_{s,s^\prime}\left(\bf{k}\right)=\epsilon _{0}\left(\bf{k}\right)+s^\prime\sqrt{\left(M^\prime_{0}-M_{1}k_{z}^{2}-M_{2}k_\parallel^2-s\lambda\right)^2+v_s^2k_\parallel^2},
\end{split}
\end{eqnarray}
where $s^\prime=+$ ($-$) refers to the conduction (valence) band. From Eqs. (\ref{m2}-\ref{m3}), one can see that there are two main effects induced by the light. One is that the parameters $C_0$ and $M_0$ are modified by the new terms $C_{2}k_{A}^{2}$ and $-M_{2}k_{A}^{2}$ respectively. The former would shift all the bands downward in the energy direction and the latter would change the positions of the Dirac points. Even so, the spin degeneracy of the system is still undisturbed. Another effect is that new spin-dependent terms $-s\lambda \tau_{z}-v_{A}\left( k_{x}\tau _{x}-sk_{y}\tau _{y}\right)$ are generated to break the TRS. In this scenario, the spin-up band $E_{+,s^\prime}$ and the spin-down band $E_{-,s^\prime}$ would exhibit different responses to $k_A$. Naturally, the spin degeneracy of the Dirac points can be destroyed.
\par
Interestingly, along with the broken spin degeneracy come various topological phases for different $k_A$, as stated in Ref. \cite{xiaoshi}. To observe the phase transitions intuitively, we plot the energy dispersions of $k_z$ axis in Fig. 2, where $\epsilon_0$ is temporarily dropped (i.e., $\epsilon_0=0$). For $k_A=0$, the bands are spin-degenerated and the system stays in the DSM phase [Fig. 2(a)]. Due to $M_{0/1}<0$, there exists two Dirac points with positions (0, 0, $\pm k_0$) ($k_0=\sqrt{M_0/M_1}$), each of which contains two spin-resolved Weyl points with opposite chiralities. Different from the paired Weyl points at the same Dirac point in conventional DSMs, the overlapped Weyl points here are nonpaired and protected from mixing by the $\mathds{Z}_2$ symmetry\cite{xiaoshi}. Considering the block-diagonal form of the Hamiltonian $H(\mathbf{k})$ [Eq. (\ref{m1})] in spin space, one can find that only the Weyl points with opposite chiralities and the same spin can form a pair, i.e., Weyl partners are locked with spin. Once a small $k_A$ is considered, i.e., $0<k_A<k_c$ with $k_c=\sqrt{M_0/(M_2- v_0^2/\hbar \Omega)}$, the two pairs of Weyl partners are separated due to the different responses of the spin-up and spin-down bands to $k_A$ [Fig. 2(b)]. Consequently, the original DSM is transformed to be a WSM, whose Weyl points are located at (0, 0, $\pm k_{0,s}$) with $k_{0,s}=\sqrt{(M_0-M_2k_A^2-s\lambda)/M_1}$. As $k_A>k_c$ is satisfied, the system enters into the WHM phase. Compared to the WSM, the key characteristic of the WHM is the fully spin-polarized property, i.e., one spin band is in the WSM phase while the other is in the insulator phase, as shown in Fig. 2(d). Due to the peculiar band structure, only the electrons of one spin band is allowed to participate in the transport when the Fermi energy $u_F$ is inserted in the gap of the other spin band. Based on this property, the WHM has been proposed as a perfect spin filter\cite{xiaoshi}. If a large $k_A$ with $k_A>\sqrt{M_0/(M_2+v_0^2/\hbar \Omega)}$ is considered, the material is changed to be a normal insulator, where all bands are gapped. In this paper, we only focus on the topological phases (DSM, WSM and WHM) since all the RKKY components vanish in the insulator phase if $u_F$ is inserted in the energy gap. Similar vanished RKKY interaction has already been discussed in the phosphorene\cite{rkky4}.
\par
In addition to the various phases mentioned above, the phase boundary ($k_A=k_c$) between the WSM and the WHM attracts us due to its peculiar dispersion, whose shape is highly spin-dependent. As shown in Fig. 2(e), the spin-up band around the Weyl points is linear in all directions. Remarkably different from this, the spin-down band is linear in $k_{x}$ (or $k_y$) axis but disperses quadratically in $k_z$ axis, i.e., exhibiting a S-DSM-type dispersion. Since the RKKY interaction is sensitive to the shape of the band structure, thus the magnetic signals characterizing the phase boundary are expected.
\par
To verify the phase transitions shown in Fig. 2, one can calculate the spin-dependent Hall conductivity $\zeta_{xy}^s$. Following the calculation processes in Refs. \cite{F8,KYYang}, one can obtain $\zeta_{xy}^s=sk_{0,s}e^2/\pi h$. In this way, one can find a net zero Hall conductivity in the DSM due to $\zeta_{xy}^+=-\zeta_{xy}^-$. Differently, a finite Hall conductivity is obtained in the WSM since $\zeta_{xy}^+$ and $\zeta_{xy}^-$ are opposite in sign and unequal in amplitude. More interestingly, a pure spin current can be induced in the WHM due to the vanished $\zeta_{xy}^-$ and survived $\zeta_{xy}^+$. As a result, different phases can be distinguished.
\par
Unlike traditional electrical methods in characterizing different phases, we attempt to build the relationship between the magnetic signals and the phase transitions in this work. To construct a model for the indirect magnetic interaction (i.e.,  the RKKY interaction), two impurities are assumed to be embedded in the bulk of the material. One impurity is located at $\mathbf{ r}_1$ and the other is at $\mathbf{ r}_2$. Each impurity would interact with the electrons of host material in a contact interaction $H_{int}=J\mathbf{S}^i\cdot\mathbf{\sigma}\delta\left(\mathbf {r}-\mathbf {r}_i\right)$, where $\mathbf{S}^i$ ($i=1$ or $2$)  denotes the spin of impurity. The two impurities would couple indirectly with each other by the itinerant electrons, thus an effective indirect exchange interaction is generated between two impurities. Using the standard perturbation theory\cite{pertub1,pertub2,pertub3,pertub4} by keeping $J$ to the second order term, the effective exchange interaction between impurities is given by
\begin{eqnarray}\label{m4}
H_{\rm R}=-\frac{J^{2}}{\pi }\mathrm{Im}\int_{-\infty }^{u_{F}}\mathrm{Tr}%
\left[ \left( \mathbf{S}^1\cdot \mathbf{\sigma }\right) G\left( \omega
,\mathbf{R}\right) \left( \mathbf{S}^2\cdot \mathbf{%
\sigma }\right) G\left( \omega ,-\mathbf{R}\right) %
\right]d\omega,
\end{eqnarray}
where zero temperature is considered and $G\left( \omega,\mathbf{R}\right)$ is the retarded Green's function with $\mathbf{R}=\mathbf{r}_{1}-\mathbf{r}_{2}$.
\par
Before evaluating the RKKY interaction, the retarded Green's functions of real space have to be derived. Using the system Hamiltonian $H(\mathbf{k})$, the retarded Green's function $G\left( \omega,\mathbf{R}\right)$ can be constructed in Lehmann's representation and it reads as
\begin{eqnarray}\label{m5}
G\left( \omega,\mathbf{R}\right)=\frac{1}{\left(2\pi\right)^3}\int e^{i\mathbf{k}\mathbf{R}}\frac{1}{\omega+i0^+-H\left(\mathbf{k}\right)}d^3\mathbf{k}.
\end{eqnarray}
Inserting the Hamiltonian $H(\mathbf{k})$ of Eq. (\ref{m1}) into the above equation, one can obtain
\begin{eqnarray}\label{m6}
G\left( \omega ,\pm\mathbf{R}\right) =\left(
                                                              \begin{array}{cc}
                                                                G_{+}\left( \omega ,\pm\mathbf{R}\right) & {\bf 0} \\
                                                                {\bf 0} & G_{-}\left( \omega ,\pm\mathbf{R}\right) \\
                                                              \end{array}
                                                            \right)
\end{eqnarray}
with
\begin{eqnarray}\label{m7}
G_{s}\left( \omega ,\pm\mathbf{R}\right)=\frac{1}{\left( 2\pi \right) ^{3}}\int d^{3}\mathbf{k}e^{i\mathbf{%
kR}}\frac{\omega_+ -\epsilon _{0}\left( \mathbf{k}\right) +\mathbf{h}_{s}\left( \mathbf{%
k}\right) \cdot \tau}{\left[ \omega_+ -\epsilon _{0}\left( \mathbf{k}\right) \right]
^{2}-h_{s}^{2}\left( \mathbf{k}\right) },
\end{eqnarray}
where $\omega_+=\omega+i0^+$ and $h_{s}\left( \mathbf{k}\right)=|\mathbf{h}_{s}\left( \mathbf{k}\right)|$. Since the Hamiltonian $H\left(\bf{k}\right)$ of Eq. (\ref{m1}) is diagonal in spin space, the retarded Green's function can also be expressed as a diagonal form in Eq. (\ref{m6}). Inserting the expressions of $ \epsilon _{0}\left( \mathbf{k}\right)$ and $\mathbf{h}_{s}\left( \mathbf{k}\right)$ [Eq. (\ref{m2})] into the above equation and integrating out the momentum $k_z$ and the angle $\varphi$ [$\tan(\varphi)=k_y/k_x$], $G_{s}\left( \omega ,\pm\mathbf{R}\right)$ ($s=\pm$) can be calculated as
\begin{eqnarray}\label{m8}
G_{s}\left( \pm \mathbf{R},\omega \right) =\begin{pmatrix}
r_{s}+t_{s} & \pm se^{is\varphi _{R}}q_{s} \\
\pm se^{-is\varphi _{R }}q_{s} & r_{s}-t_{s}%
\end{pmatrix},
\end{eqnarray}
where $\varphi _{R}=\arctan(R_y/R_x)$. $r_{s}$, $t_{s}$ and $q_{s}$ are given by
\begin{eqnarray}\label{m9}
\begin{split}
r_{s}=&\int_{0}^{\infty }\left( \omega ^{\prime }f_{s}+C_{1}\frac{d^{2}f_{s}%
}{dR_{z}^{2}}\right) \frac{k_{\parallel }J_{0}\left( k_{\parallel
}R_{\parallel }\right) }{4\pi \left( C_{1}^{2}-M_{1}^{2}\right) \left(
g_{s,+}-g_{s,-}\right) }dk_{\parallel }, \\
t_{s}=&\int_{0}^{\infty }\left( M_{s}f_{s}+M_{1}\frac{d^{2}f_{s}}{dR_{z}^{2}}%
\right) \frac{k_{\parallel }J_{0}\left( k_{\parallel }R_{\parallel }\right)
}{4\pi \left( C_{1}^{2}-M_{1}^{2}\right) \left( g_{s,+}-g_{s,-}\right) }%
dk_{\parallel }, \\
q_{s}=&\int_{0}^{\infty }\frac{iv_{s}f_{s}k_{\parallel }^{2}J_{1}\left(
k_{\parallel }R_{\parallel }\right) }{4\pi \left( C_{1}^{2}-M_{1}^{2}\right)
\left( g_{s,+}-g_{s,-}\right) }dk_{\parallel },
\end{split}
\end{eqnarray}
where
\begin{eqnarray}\label{m10}
\begin{split}
f_{s}&=\frac{e^{-\sqrt{g_{s,-}}R_{z}}}{\sqrt{g_{s,-}}}-\frac{e^{-\sqrt{%
g_{s,+}}R_{z}}}{\sqrt{g_{s,+}}}, \\
\omega ^{\prime }&=\omega _{+}-C_{0}^{\prime }-C_{2}k_{\parallel
}^{2}, \\
M_{s}&=M_{0}^{\prime }-M_{2}k_{\parallel }^{2}-s\lambda, \\
g_{s,\pm }&=\frac{M_{s}M_{1}-\omega ^{\prime }C_{1}}{C_{1}^{2}-M_{1}^{2}}\pm
\sqrt{\frac{\left( \omega ^{\prime }M_{1}-M_{s}C_{1}\right) ^{2}}{\left(
C_{1}^{2}-M_{1}^{2}\right) ^{2}}+\frac{v_{s}^{2}k_{\parallel }^{2}}{%
C_{1}^{2}-M_{1}^{2}}}.
\end{split}
\end{eqnarray}
In Eq. (\ref{m9}), $J_{v}(x)$ is the $n$th order Bessel function of the first kind.
\par
Inserting the Eqs. (\ref{m6}) and (\ref{m8}) into the Eq. (\ref{m4}) and cancelling the spin and orbital degrees of freedom, the RKKY interaction $H_{\rm R}$ can be split into the following components,
\begin{eqnarray}\label{m11}
H_{\rm R}=\sum_{i=x,y}J_{ii}{S^{1}_{i}}{S^{2}_{i}}+J_{zz}{S^{1}_{z}}{S^{2}_{z}}
\end{eqnarray}
with
\begin{eqnarray}\label{m12}
\begin{split}
J_{xx,yy}&=\frac{-4J^{2}}{\pi }{\rm{Im}}\int_{-\infty }^{u_{F}}\left[
r_{+}r_{-}+t_{+}t_{-}+q_{-}q_{+}\cos \left( 2\varphi _{R}\right) \right] d\omega, \\
J_{zz}&=\frac{-2J^{2}}{\pi }{\rm{Im}}\int_{-\infty }^{u_{F}}\left(
t_{+}^{2}+r_{+}^{2}-q_{+}^{2}+t_{-}^{2}+r_{-}^{2}-q_{-}^{2}\right) d\omega.
\end{split}
\end{eqnarray}
Noting that the subscript $s=\pm$ for $r_{s}$, $t_s$ and $q_s$ refers to the spin of the electrons. Thus, one can find that $J_{xx}$ is induced by the interplay between spin-up and spin-down bands while $J_{zz}$ stems from the contribution of bands with the same spin. Due to the protected inversion symmetry of the Hamiltonian $H({\mathbf{k}})$, no DM term arises in Eq. (\ref{m11}), the same result is also found in Ref. \cite{DSMrkky1}. Since $J_{xx}=J_{yy}$ always stands, the RKKY interaction of Eq. (\ref{m11}) can be expressed in another form, which is given by
\begin{eqnarray}\label{m13}
H_{\rm R}=J_{H}\mathbf{S}^1\cdot \mathbf{S}^2+J_{I}S^1_zS^2_z
\end{eqnarray}
with
\begin{eqnarray}\label{m14}
\begin{split}
J_{H}&=J_{xx},  \\
J_{I}&=J_{zz}-J_{xx},
\end{split}
\end{eqnarray}
where $J_{H}$ is the Heisenberg term and $J_{I}$ is the Ising term.
\par
\section{RKKY signals characterizing the topological phases in the absence of $\epsilon_0(\mathbf{k})$}
 \begin{figure}[!htb]
\centering \includegraphics[width=0.39\textwidth]{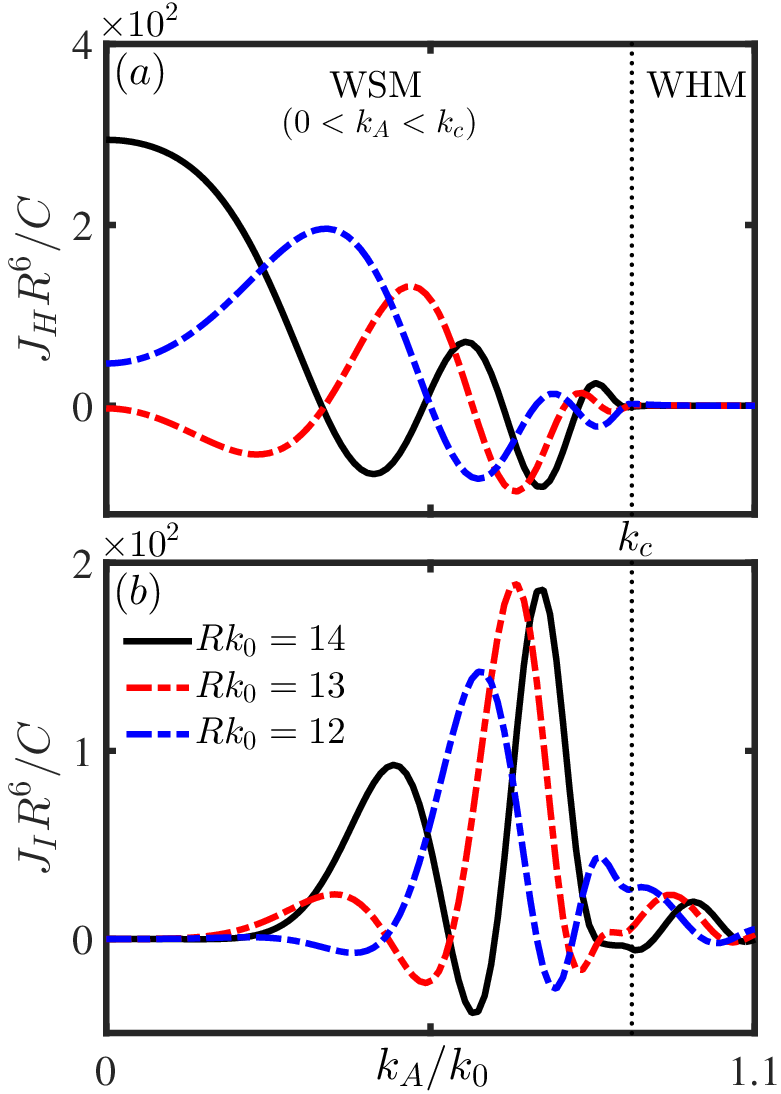}
\caption{(Color online) The RKKY components (a) $J_H$ and (b) $J_I$ versus the light intensity $k_A$ with different impurity distances. Impurities are deposited on the $z$ axis, and $u_F = 0$ is considered. The vertical dotted lines denote the phase boundary ($k_A=k_c$) between the WSM and the WHM.}
\end{figure}
In this section, we calculate the RKKY interaction in the absence of the term $\epsilon_0(\mathbf{k})$. The effect of $\epsilon_0(\mathbf{k})$ on the RKKY interaction would be discussed in the next section.

\subsection{The case with impurities deposited on $z$ axis}
In this subsection, we focus on the case with impurities deposited on $z$ axis (i.e., the line connecting the Weyl points). In this impurity configuration, one can find $q_\pm=0$ by checking the Eq. (\ref{m9}), where $J_1\left(k_\parallel R_\parallel\right)=0$ for the case of $R_\parallel=0$. Plugging $r_s$ and $t_s$ [Eqs. (\ref{m9}-\ref{m10})] into the Eq. (\ref{m12}), the RKKY interaction of zero Fermi energy ($u_F=0$) can be calculated numerically. To carry out the discussion, the RKKY components $J_H$ and $J_I$ versus the light intensity $k_A$ are plotted. According to the Fig. 3, it is found that the RKKY components exhibit significantly different behaviors in different phases, as described blow:
\begin{enumerate}
 \item In the DSM ($k_A=0$), there only exists the Heisenberg term $J_H$, which supports an isotropic XXX ($J_{xx}=J_{yy}=J_{zz}$) spin model for the impurities. The same result has also been reported in Refs. \cite{DSMrkky1,DSMrkky2}. For the vanished $J_I$ in the DSM, the related mechanism is attributed to the spin degeneracy under the protection of the TRS. It can be further understood by checking the Eqs. (\ref{m9}-\ref{m10}) and (\ref{m12}), where $r_+=r_-$ and $t_+=t_-$ (the subscripts $\pm$ for spin up and down) lead to $J_{xx,yy}=J_{zz}$;
\item In the WSM ($0<k_A<k_c$), a nonzero Ising term $J_I$ arises ($J_I=J_{zz}-J_{xx}\neq 0$) due to the broken spin degeneracy. In this scenario, $J_I$ coexists with $J_H$. As a result, an anisotropic $XXZ$ spin model ($J_{xx}=J_{yy}\neq J_{zz}$) is generated, which is different from the original isotropic spin model in the DSM;
\item For the WHM ($k_A>k_c$), it acts like a filter, i.e., $J_I$ still survives but $J_H$ is filtered out once the system is transformed from the WSM to the WHM. As far as we know, this magnetic filtering effect is unique to the WHM, which distinguishes the WHM from the other 3D materials. Physically, this effect is a reflection of the fully spin-polarized property (i.e., one spin band is in the insulator phase while the other is in the WSM phase). More specifically, the gapped band leads to the vanished $J_H$ while finite $J_I$ is mainly contributed by the WSM-type band. Further explanations are organized as:
    \begin{enumerate}
    \item To explain the vanished $J_H$ in the WHM, one have to study the decaying law of $J_H$, since the amplitude of the interaction is mainly determined by the decaying law. We plot the $R$-dependent $J_H$ with different $k_A$ in Fig. 4. In the WHM [Fig. 4(c)], there arises an exponential decaying law $J_H\propto e^{-\kappa^0_zR}$ ($\kappa^0_z=\sqrt{-E_g/2M_1}$) , which decays faster than the cases in the WSM ($k_A<k_c$) and at the phase boundary ($k_A=k_c$). To understand this exponential law, one have to notice that $J_H$ is contributed by the terms $r_+r_-$ and $t_+t_-$ in Eq. (\ref{m12}). Here, $r_-$ and $t_-$ are induced by the spin-down band, whose energy gap $E_g$ makes the solution of $k_z$ to be an imaginary number (i.e., $k_z=i\kappa^0_z$). Thus, $J_H(k_A>k_c)\propto e^{-\kappa^0_zR}$ is obtained since $k_z$ couples with $R$ in the phase factor $e^{ik_zR}$ of the Green's function [Eq. (\ref{m5})]. Due to the large value of $\kappa^0_z$, $J_H\propto e^{-\kappa^0_zR}$ vanishes if the long-range case (i.e., relatively large $R$) is considered. For example, $J_H$ almost vanishes for $Rk_0>10$ in Fig. 4(c). This explains the vanished $J_H$ in the WHM shown in Fig. 3(a).

    \item Different from $J_H$, $J_I=J_{zz}-J_{xx}$ is mainly contributed by the terms $t_+^2$ and $r_+^2$ in Eq. (\ref{m12}), since the other terms related to $r_-$ (or $t_-$) can be ignored due to the exponential decaying law (as stated in previous paragraph). Noting that $t_+^2$ and $r_+^2$ are completely induced by the spin-up WSM-type band, which usually generates a decaying law of $J_I\propto1/R^5$, similar to the case of Fig. 4(a). Thus, the amplitude of $J_I$ is still considerable in the WHM phase [Fig. 3(a)].

    \end{enumerate}
\end{enumerate}
\par
\begin{figure}[!htb]
\centering \includegraphics[width=0.39\textwidth]{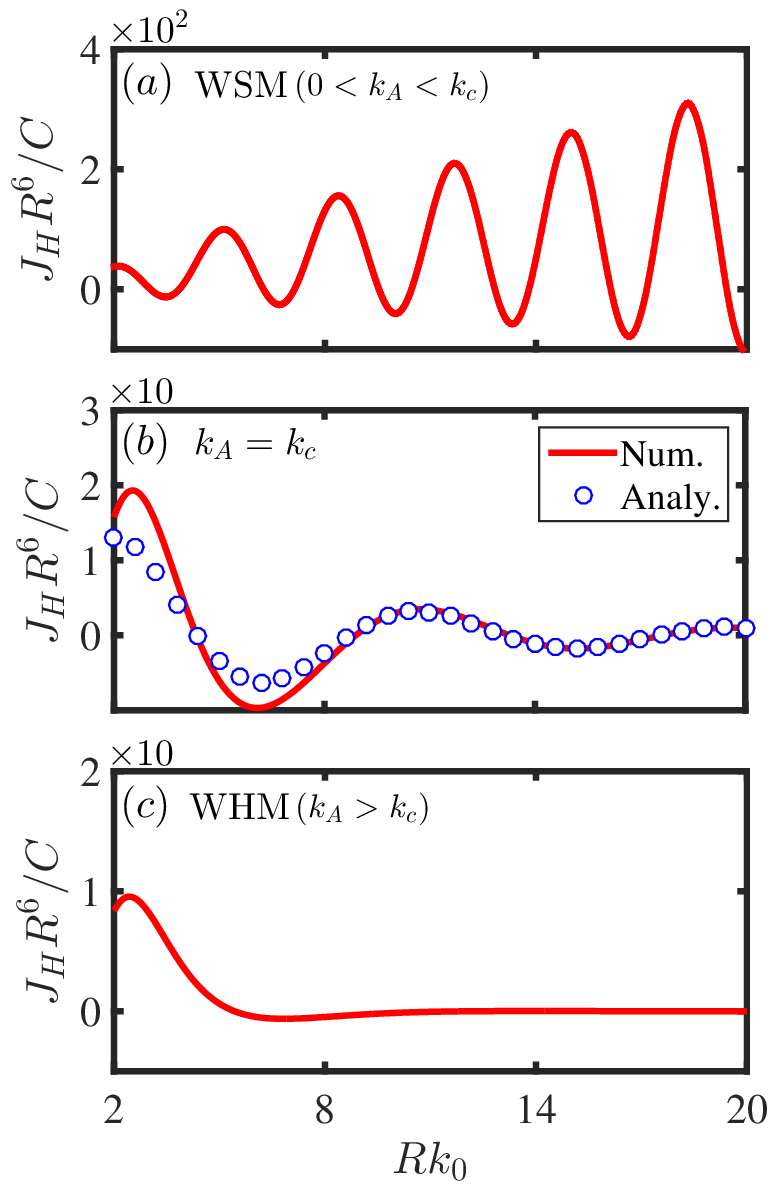}
\caption{(Color online)  Spatial dependence of $J_{H}$ with (a) $k_A=0.37k_c$ in the WSM ($0<k_A<k_c$), (b) $k_A=k_c$ and (c) $k_A=1.18k_c$ in the WHM. The hollow circles in (b) denote the analytical result of the Eq. (\ref{II10}) in the Appendix II, and the solid lines in (a-c) refer to the numerical results calculated from Eqs. (\ref{m9}-\ref{m10},\ref{m12},\ref{m14}).}
\end{figure}
In a brief summary, we obtain $J_H\neq0$, $J_I=0$ for the DSM, $J_{H,I}\neq0$ for the WSM, and $J_{H}=0$, $J_I\neq0$ for the WHM. Thus, one can identify the phase transition of DSM/WSM by checking the Ising term $J_I$, and the Heisenberg term $J_H$ can be used to distinguish the WSM from the WHM.

\subsection{The case with impurities placed in a direction deviated from $z$ axis.}
\begin{figure}[!htb]
\centering \includegraphics[width=0.44\textwidth]{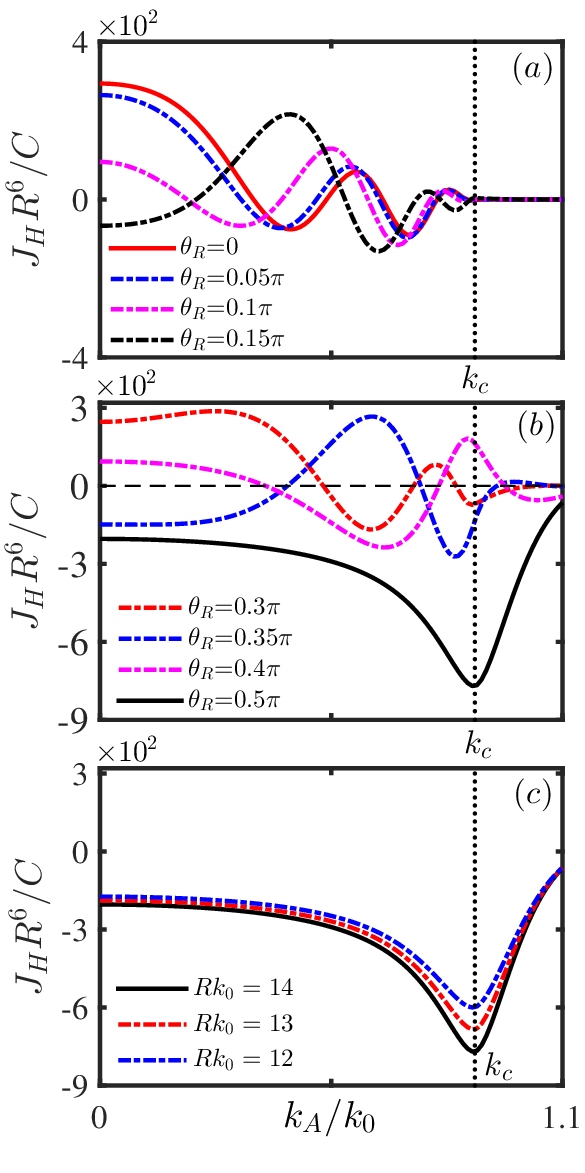}
\caption{(Color online) (a-c) The Heisenberg term $J_H$ versus the light intensity $k_A$ with $u_F=0$ and $\varphi_R=\pi/4$. Here, the vertical dotted lines denote the phase boundary ($k_A= k_c$) between the WSM and the WHM. (a, b) Different polarization angles $\theta_R$ are considered with $Rk_0=14$. (c) Impurities are placed in $x$-$y$ plane (i.e., $\theta_R=\pi/2$) with different impurity distances $R$.}
\end{figure}

\begin{figure}[!htb]
\centering \includegraphics[width=0.41\textwidth]{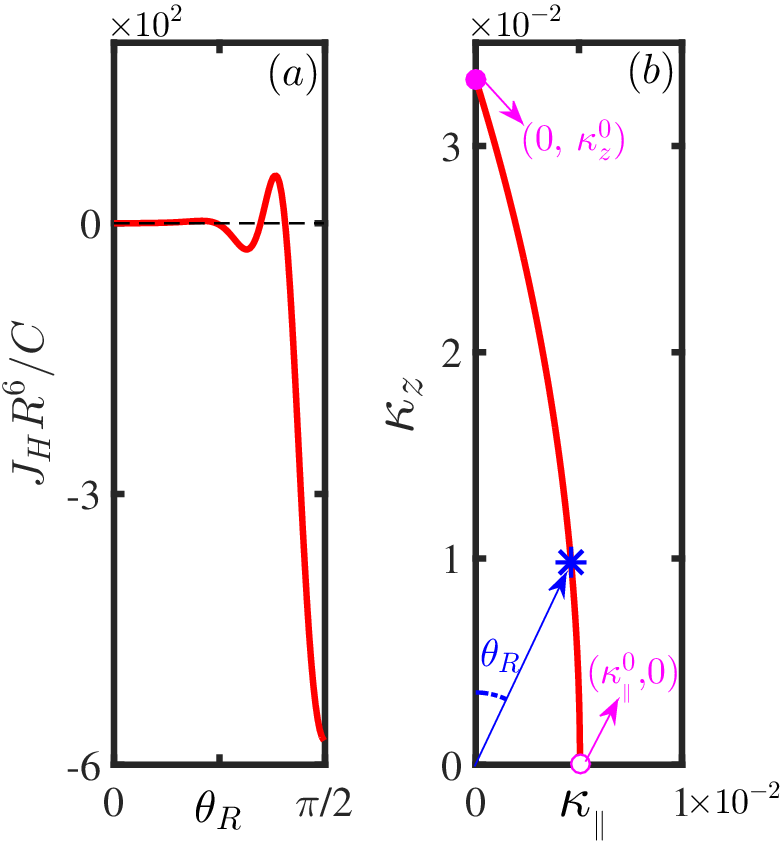}
\caption{(Color online) (a) $\theta_R$-dependent $J_H$ in the WHM phase ($k_A=0.95k_0$) with $\varphi_R=\pi/4$ and $Rk_0=14$. (b) The relationship between $\kappa_z$ and $\kappa_\parallel$. Here, the coordinate system of real space is consistent with that of $\mathbf{k}$ space. The asterisk denotes the value of $\kappa=\sqrt{\kappa_z^2+\kappa_\parallel^2}$ taken in the direction of $\theta_R$.}
\end{figure}

\begin{figure}[!htb]
\centering \includegraphics[width=0.39\textwidth]{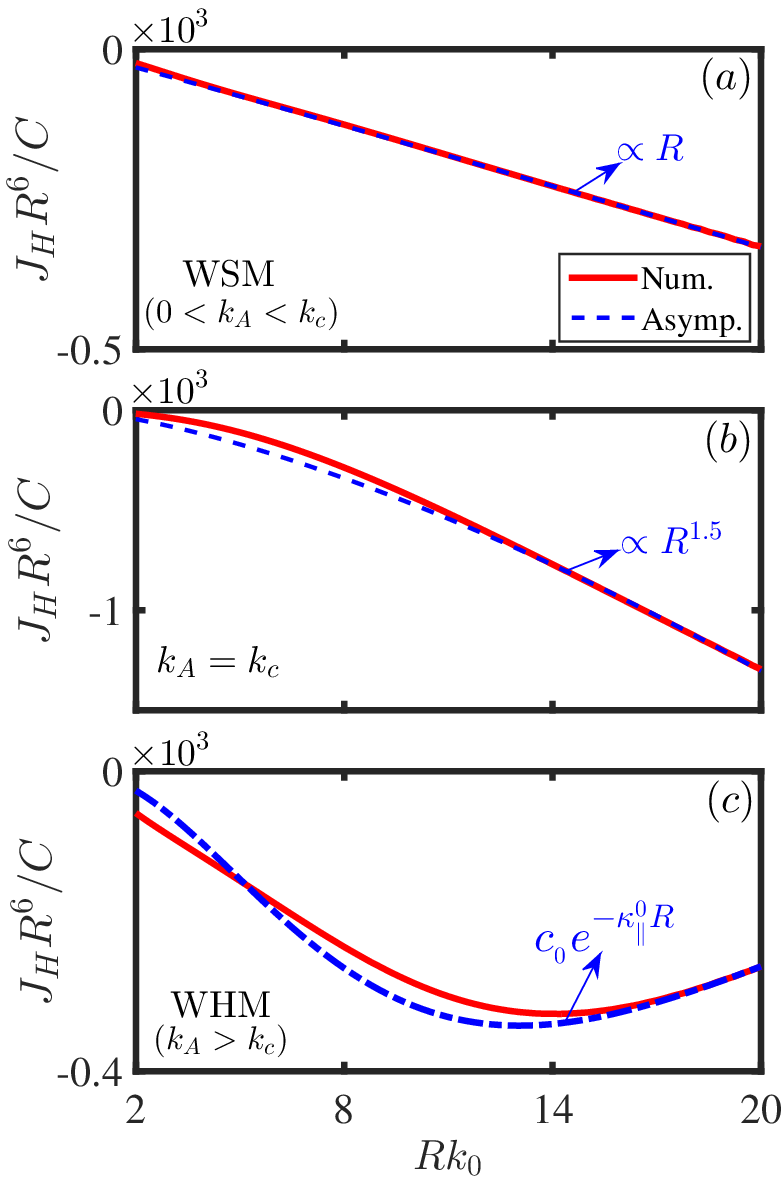}
\caption{(Color online) (a-d) Spatial dependence of the Heisenberg term $J_H$  with (a) $k_A=0.37k_c$ in the WSM ($0<k_A<k_c$), (b) $k_A=k_c$ and (c) $k_A=1.24k_c$ in the WHM. The dashed lines denote the long-range asymptotic results for $J_H$. Here, $c_{_0}=-0.0165R^{-3.5}$, $\kappa_\parallel^0=E_g/2v_-$ and $\varphi_R=\pi/4$.}
\end{figure}
In practice, due to the limitation in the accuracy of the doping techniques, impurities can not be precisely placed in the $z$ axis. One may wonder whether the magnetic signals of Fig. 3(a) still exist when impurities are placed in a direction deviated from $z$ axis. The polarization angle $\theta_R$ [$\tan(\theta_R)=R_\parallel/R_z$] is used to evaluate the deviation between the direction of the impurities and the $z$ axis. In the following, we only focus on the Heisenberg term $J_H$ since the signal for the phase transition of DSM/WSM carried by $J_I$ is independent on $\theta_R$.
\par
For a small angle $\theta_R$ ($\theta_R\leq0.15\pi$), it is found that $J_H$ still can be used to distinguish the WHM from the WSM, as shown in Fig. 5(a), where the results are similar to the case in Fig. 3(a). If a relatively large $\theta_R$ is considered (except $\theta_R=\pi/2$), $J_H$ is failed in characterizing the phase transition of WSM/WHM, as indicated by the dashed lines in Fig. 5(b). The reason is that the original vanished $J_H$ is changed to be a finite one if $\theta_R$ increases substantially, as shown in Fig. 6(a). The high dependence of $J_H$ on $\theta_R$ is attributed to the anisotropic imaginary wave number $k=\sqrt{k_z^2+k_\parallel^2}=i\sqrt{\kappa_z^2+\kappa_\parallel^2}=i\kappa$, which is induced by the anisotropic gapped spin-down band $E_{-,s^\prime}$. Similar to the case of Fig. 4(c), an exponential decay $J_H\propto e^{-\kappa R}$ is generated. Using the equation $E_{-,s^\prime}=u_F=0$, the relationship between $\kappa_z$ and $\kappa_\parallel$ is plotted in Fig. 6(b), where the coordinate system of real space is consistent with that of $\mathbf{k}$ space. In order to make an effective contribution to the RKKY interaction, the value of $\kappa$ must be taken in the direction of $\theta_R$, as highlighted by an asterisk in Fig. 6(b). According to the Fig. 6(b), one can find $\kappa=\sqrt{(\kappa_z^0)^2+0^2}=\kappa_z^0$ for $\theta_R=0$. In this case, $J_H(\theta_R=0)$ recovers the result of Fig. 4(c), i.e., $J_H(\theta_R=0)\propto e^{-\kappa_z^0 R}$. As $\theta_R$ changes from $0$ to $\pi/2$ (i.e., impurities are moved from $z$ axis to $x$-$y$ plane), $\kappa$ decreases dramatically, which brings a great enhancement to the amplitude of $J_H$ because of $J_H\propto e^{-\kappa R}$ ($R>0$).
\par
Although the signal of $J_H$ similar to that of Fig. 3(a) is destroyed by the large $\theta_R$, there emerges another signal for $J_H$ to characterize the phase transition of WSM/WHM once the impurities are moved into the $x$-$y$ plane (i.e., $\theta_R=\pi/2$), as indicated in Figs. 5(b) and 5(c). In the vicinity of the critical point $k_A=k_c$, $J_H$ in the WSM increases with $k_A$, but it decreases with $k_A$ in the WHM. As a consequence, there emerges a significant dip exactly at $k_A=k_c$, which provides an unambiguous signal to ascertain the phase boundary between the WSM and the WHM. The signal here can also be understood by checking the decaying laws of $J_H$. According to the long-range asymptotic behaviors of $J_H$ in Fig. 7, one can find a slowest decaying law $J_H\propto 1/R^{4.5}$ at $k_A=k_c$ as compared to the cases in the WSM and the WHM. This law is a result of the interplay between the spin-up Weyl band and the spin-down S-DSM-type band. Noting that the Weyl band contributes a decaying law of $1/R^5$ for the interaction while the S-DSM-type band induces a slowly decaying law of $1/R^4$\cite{semiDirac}. As a result, the interlay of these two bands generates an intermediate decaying law $J_H\propto 1/R^{4.5}$, which is further verified by the analytical result of Eq. (\ref{II19}) in the Appendix II. Due to the slowest decaying law, the largest amplitude of $J_H$ is naturally generated at the phase boundary ($k_A=k_c$). This explains the dip structure shown in Fig. 5(c).
\begin{figure}[!htb]
\centering \includegraphics[width=0.39\textwidth]{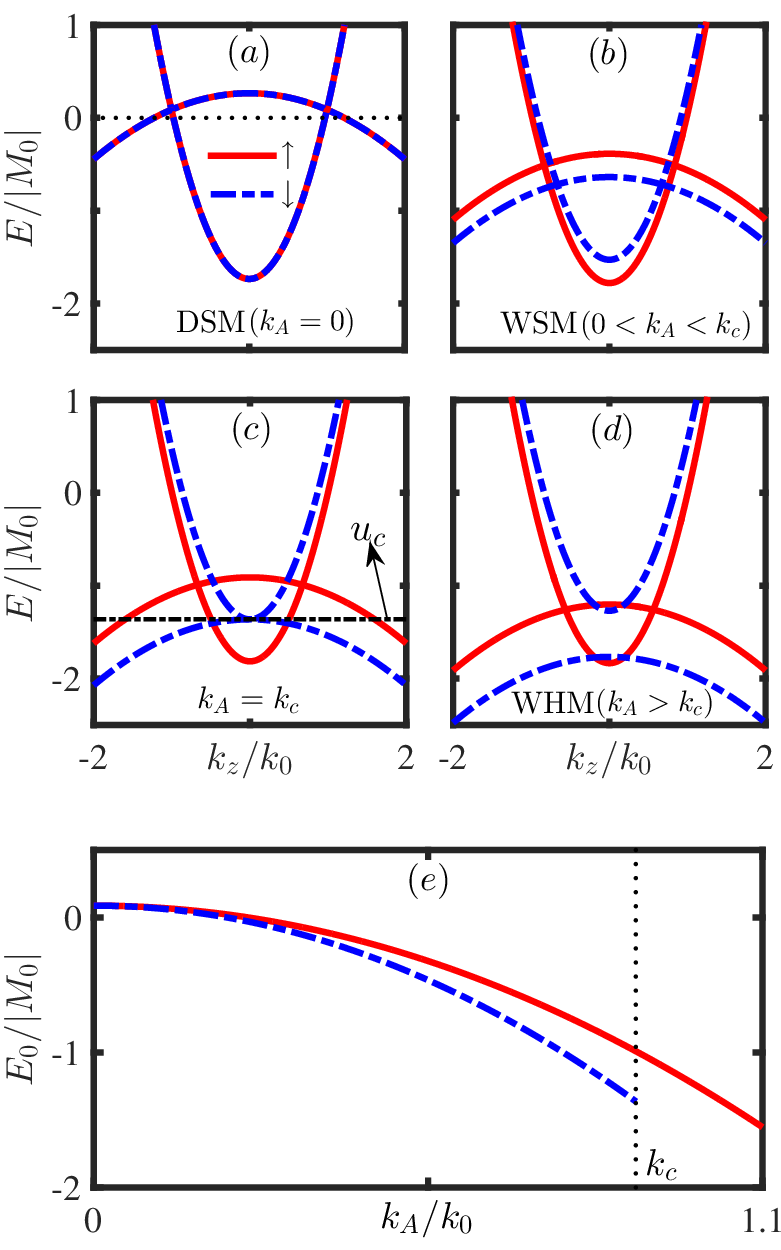}
\caption{(Color online) (a-d) Energy dispersion along the $k_z$ axis for different phases in the presence of $\epsilon_0(\mathbf{k})$. Other parameters are the same as that in Figs. 2(a-d). $u_c$ in (c) refers to the specific Fermi energy at which the spin-down conduction band touches the valence band. (e) The energy of the Weyl points of different spins versus $k_A$. Parameters $C_0=-0.06382$ ${\rm eV}$, $C_1=8.7536$ ${\rm eV}$ ${\rm \AA^2}$, $C_2=-8.4008$ ${\rm eV}$ ${\rm \AA^2}$ are extracted from ${\rm Na_3Bi}$\cite{ZWang1} material.}
\end{figure}
\section{The effect of $\epsilon_0(\mathbf{k})$ on the RKKY signals}
\begin{figure}[!htb]
\centering \includegraphics[width=0.46\textwidth]{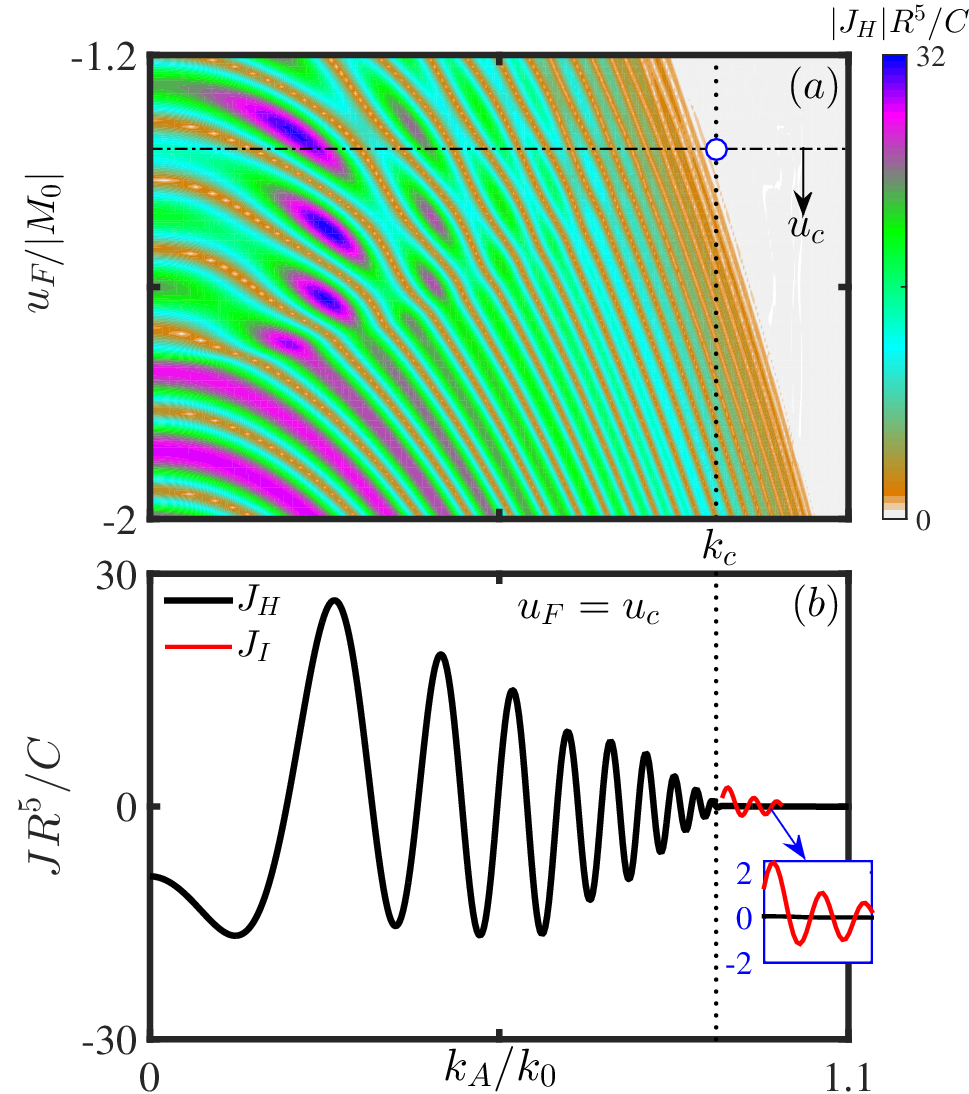}
\caption{(Color online) (a) The Heisenberg term $J_H$ as a function of $k_A$ and $u_F$. (b) $k_A$-dependent RKKY components $J_H$ and $J_I$ with $u_F=u_c$. $u_c$ is the specific Fermi energy as depicted in Fig. 8(c), and the red solid line (enlarged in the illustration) refers to the Ising term $J_I$ for a small interval of $k_A$ in the WHM. All results in (a-b) are calculated by considering the effect of $\epsilon_0$, and impurities are deposited on $z$ axis with $Rk_0=14$.}
\end{figure}
\begin{figure}[!htb]
\centering \includegraphics[width=0.39\textwidth]{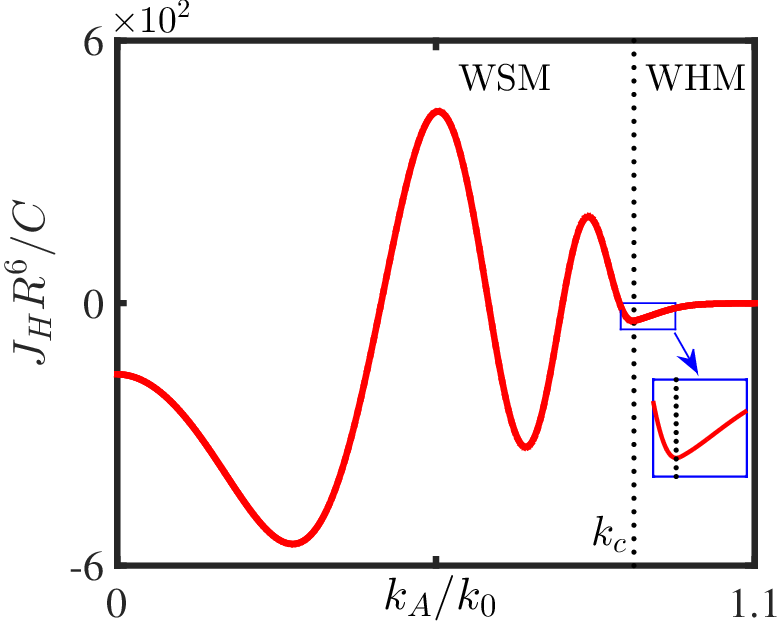}
\caption{(Color online) The Heisenberg term $J_H$ as a function of $k_A$ with $u_F=u_c$ and $Rk_0=14$. Here, impurities are placed in $x$-$y$ plane with $\varphi_R=\pi/4$. The vertical dotted lines denote the phase boundary ($k_A= k_c$) between the WSM and the WHM. }
\end{figure}
In this section, the effect of $\epsilon_0(\mathbf{k})$ on the RKKY interaction would be discussed. Our ultimate purpose is to test whether the magnetic signals characterizing the phase transitions, as well as the signal for the fully spin-polarized property of the WHM, are still valid.
\par
Before exploring the magnetic signals, we would briefly discuss the effect of $\epsilon_0(\mathbf{k})$ on the band structure, which is plotted in Fig. 8. First, we consider the case of DSM with $k_A=0$. Due to the broken electron-hole symmetry, $\epsilon_0(\mathbf{k})$ would bring three main effects [Fig. 8(a)], which are organized as: (1) The low-energy bands around the Dirac points are slightly tilted. (2) An asymmetry between the conduction and valence bands arises, and along with the deformed Fermi surface. (3) The energy bands are lifted as a whole. Once the light intensity $k_A$ is turned on, the effect of (3) would be substantially modified. As shown in Figs. 8(b-d), the larger $k_A$ is, the more pronounced is the movement of all energy bands in the negative-energy direction. This movement can also be seen in Fig. 8(e), where the energies of the Weyl points of different spins versus $k_A$ are plotted. Although significant modifications are induced for the energy bands by $\epsilon_0(\mathbf{k})$, the topological phase transitions are undisturbed. Thus, magnetic signals exhibited in previous section are expected to be preserved.
\par
In order to find the survived magnetic signals, one have to display the numerical results of $J_H$ as a function of the Fermi energy $u_F$, since the bands are drastically shifted in the energy direction as $k_A$ varies. First, the case with impurities deposited on $z$ axis is considered. In Fig. 9(a), we plot $|J_H|$ as a function of $u_F$ and $k_A$. One can find that the existence of the magnetic signal at $k_A=k_c$ is highly dependent on the selection of the Fermi energy. The signal [the circle in Fig. 9(a)] identifying the phase boundary between the WSM and the WHM can be obtained only in the condition of $u_F=u_c$, which corresponds precisely to the energy of the closing point of the spin-down bands [Fig. 8(c)]. The signal here is similar to that of Fig. 3(a), as verified by the $k_A$-dependent $J_H$ with the Fermi energy $u_c$ in Fig. 9(b). The survival of the magnetic signal can be understood by reviewing the three effects stated in the previous paragraph. By choosing $u_F=u_c$, the amplitude of $J_H$ around $k_A=k_c$ is only modified by the tilting effect of the band while the other two effects can be ignored. Noting that the magnetic signal here is mainly determined by the decaying laws of the interaction, which can not be changed by the tilting effect of the band\cite{Weyl3}. Thus, the magnetic signal still survives in the presence of $\epsilon_0(\mathbf{k})$. In addition, the signal characterizing the fully spin-polarized property also survives, as indicated by the illustration of Fig. 9(b), where $J_H=0$ and $J_I\neq0$ in the WHM. Similarly, we plot the $k_A$-dependent $J_H$ with impurities in $x$-$y$ plane in Fig. 10, where the Fermi energy is also set as $u_F=u_c$. One can find that there still exists a dip structure for $J_H$ at the phase boundary ($k_A=k_c$).

\section{Summary}
We have explored the RKKY interaction in ${\rm Na_3Bi}$-type DSMs subject to an off-resonant light, which can change the original DSM to the WSM and even to the WHM. It is found that signals can be extracted from the RKKY interaction for characterizing the topological phase transitions. For the phase transition of DSM/WSM, it can be identified by the Ising term $J_I$, whose existence depends on whether the spin degeneracy of the system is preserved or not. By detecting the Heisenberg term $J_H$ with impurities in the $z$ axis, the phase boundary between the WSM and the WHM can be easily ascertained. In addition, we find that only the Ising term survives in the WHM, which is a reflection of the fully spin-polarized property. For the case with impurities deposited on the $x$-$y$ plane, the dip structure of $J_H$ can also be used to identify the phase transition of WSM/WHM. Furthermore, we have proved that all magnetic signals are robust to the term that breaks the electron-hole symmetry. Our work has shown that measurement on the RKKY interaction could provide us an alternative method to probe the rich topological phases in 3D Floquet DSMs. Our proposal is feasible with the present techniques, e.g., spin-polarized scanning tunneling spectroscopy\cite{Laplane}, which can measure the magnetization curves of individual atoms, or the electron-spin-resonance technique coupled with an optical detection scheme\cite{Wiebe1,Wiebe2}. Our results also suggest that the Floquet DSMs are powerful platforms for controlling the magnetic interaction.

\acknowledgements This work was supported by the National Natural Science Foundation of China (Grant Nos. 12104167, 12174121, 11904107, 11774100), by the Guangdong Basic and Applied Basic Research Foundation under Grant No. 2023B1515020050, by GDUPS (2017) and by Key Program for Guangdong NSF of China (Grant No. 2017B030311003).
\par
S.-M. Cai, X. Wei and Y.-C. Chen contributed equally to this work.
\newpage
\begin{widetext}
\appendix
\renewcommand\thesection{\Roman{section}}
\renewcommand\thesubsection{\Alph{subsection}}
\def\CTeXPreproc{Created by ctex v0.2.12, don't edit!}
\numberwithin{equation}{section}
\section{Phase transitions induced by the off-resonant CPL in ${\rm Na_3Bi }$-type materials}
The model employed in Eq. (\ref{m1}) can be realized by considering the effect of a periodic driving to the following model of DSMs,
\begin{equation}
\label{I1}
H_{0}=C_{0}+C_{1}k_{z}^{2}+C_{2}k_\parallel^2+\left( M_{0}-M_{1}k_{z}^{2}-M_{2}k_\parallel^2 \right) \tau _{z}+v_{0}\left( k_{x}\sigma
_{z}\tau _{x}-k_{y}\tau _{y}\right),
\end{equation}
where $k_\parallel^2= k_{x}^{2}+k_{y}^{2}$, $\mathbf{\tau}=(\tau_x,\tau_y,\tau_z)$ is the vector of pauli matrix in orbital space, and the subscript $s=+$ ($-$) for spin up (down). The above model is extracted from the ${\rm Na_3B_i }$-type DSMs\cite{ZWang1,ZWang2}. For the sake of concreteness, a beam of CPL is assumed to be injected in the $z$ axis. The corresponding vector potential is described as $\mathbf{A}(t)=A_0[\cos(\Omega t),\sin(\Omega t),0]$ with period $T=2\pi/\Omega$.  By applying the Peierls substitution $\mathbf{k}\rightarrow \mathbf{k}+e\mathbf{A}/\hbar$, the system Hamiltonian becomes time-dependent. Using the Floquet theory\cite{F6} with the off-resonant condition of $\hbar\Omega\gg BW$ ($BW$ is the bandwidth), the modified part of the Hamiltonian induced by
light reads as
\begin{equation}
\label{I2}
H=V_{0}+\sum_{n\geq 1}\frac{\left[ V_{+n},V_{-n}\right] }{n\hbar
\Omega}+O\left(\frac{1}{\Omega ^{2}}\right),
\end{equation}
where $V_{n}=\frac{1}{T}%
\int_{0}^{T}H_0(\mathbf{k}+e\mathbf{A}/\hbar)e^{-in\hbar \Omega t}dt$. Specifically, $V_0$ can be calculated as
\begin{eqnarray}
\label{I3}
\begin{split}
V_0=&\frac{1}{T}\int_{0}^{T}H_{0}\left( \mathbf{k}+e\mathbf{A}/\hbar
\right) dt, \\
=&\frac{1}{T}\int_{0}^{T}\left\{ C_{0}+C_{1}k_{z}^{2}+C_{2}\left[
k_{x}+eA_{0}\cos \left( \Omega t\right) /\hbar \right] ^{2}+C_{2}\left[
k_{y}+eA_{0}\sin \left( \Omega t\right) /\hbar \right] ^{2}\right\} dt \\
&+\frac{1}{T}\int_{0}^{T}\left\{ M_{0}-M_{1}k_{z}^{2}-M_{2}\left[
k_{x}+eA_{0}\cos \left( \Omega t\right) /\hbar \right] ^{2}-M_{2}\left[
k_{y}+eA_{0}\sin \left( \Omega t\right) /\hbar \right] ^{2}\right\} \tau
_{z}dt\\
&+\frac{1}{T}\int_{0}^{T}v_{0}\left\{ \left[ k_{x}+eA_{0}\cos \left( \Omega
t\right) /\hbar \right] \sigma _{z}\tau _{x}-\left[ k_{y}+eA_{0}\sin \left(
\Omega t\right) /\hbar \right] \tau _{y}\right\} dt.
\end{split}
\end{eqnarray}
Noting that $\int_{0}^{T}\sin(\Omega t)dt=0$ and $\int_{0}^{T}\cos(\Omega t)dt=0$, thus the above equation can be further simplified as
\begin{eqnarray}
\label{I4}
\begin{split}
V_0=&\frac{1}{T}\int_{0}^{T}\left[C_{0}+C_{1}k_{z}^{2}+C_{2}k_\parallel^2+\left(
M_{0}-M_{1}k_{z}^{2}-M_{2}k_{\parallel }^{2}\right) \tau _{z}+v_{0}\left(
k_{x}\sigma _{z}\tau _{x}-k_{y}\tau _{y}\right) \right] dt\\
&+\frac{1}{T}\int_{0}^{T}\left\{ C_{2}\left[ \frac{e^{2}A_{0}^{2}\cos
^{2}\left( \Omega t\right) }{\hbar ^{2}}+\frac{e^{2}A_{0}^{2}\sin ^{2}\left(
\Omega t\right) }{\hbar ^{2}}\right] -M_{2}\left[ \frac{e^{2}A_{0}^{2}\cos
^{2}\left( \Omega t\right) }{\hbar ^{2}}+\frac{e^{2}A_{0}^{2}\sin ^{2}\left(
\Omega t\right) }{\hbar ^{2}}\right] \tau _{z}\right\} dt, \\
=&H_{0}+C_{2}k_{A}^{2}-M_{2}k_{A}^{2}\tau _{z}, \\
\end{split}
\end{eqnarray}
where $k_A=eA_0/\hbar$. As shown above, besides $H_{0}$, extra terms $C_2k_A^2$ and $-M_2k_A^2\tau_z$ are generated for $V_0$. The same extra terms are also found in Refs.\cite{xiaoshi,F8}. Similarly, one can obtain other Floquet sidebands as
\begin{eqnarray}
\label{I5}
\begin{split}
V_{\pm 1}=\frac{v_{0}\left( \sigma _{z}\tau _{x}\mp i\tau
_{y}\right) +2\left( k_{x}\pm ik_{y}\right) \left( C_{2}-M_{2}\tau
_{z}\right) }{2}k_{A},
\end{split}
\end{eqnarray}
and $V_n=0$ for $|n|\geq2$. Substituting the Eqs. (\ref{I4}-\ref{I5}) into the Eq. (\ref{I2}), one can obtain the following effective
Hamiltonian as
\begin{equation}
\label{I6}
H=H_{0}+C_{2}k_{A}^{2}-M_{2}k_{A}^{2}\tau _{z}-\lambda\sigma _{z}\tau _{z}-v_A%
\left( k_{x}\tau _{x}-k_{y}\sigma _{z}\tau _{y}\right).
\end{equation}
were $\lambda=v_{0}^{2}k_{A}^{2}/\left(
\hbar \Omega \right)$ and $v_A=2v_{0}k_{A}^{2}M_{2}/\left(\hbar \Omega \right)$.
\par
In spin space, the effective Hamiltonian of Eq. (\ref{I6}) can be rewritten as
\begin{equation}
\label{I7}
H=\left(
    \begin{array}{cc}
     \epsilon _{0}\left(\mathbf{k}\right) + \mathbf{h}_+\left(\mathbf{k}\right) \cdot\mathbf{\tau}&0 \\
      0&\epsilon _{0}\left(\mathbf{k}\right)+ \mathbf{h}_-\left(\mathbf{k}\right) \cdot\mathbf{\tau} \\
    \end{array}
  \right),
\end{equation}
with
\begin{eqnarray}
\begin{split}
\epsilon _{0}\left(\mathbf{k}\right) =&C_{0}+C_{2}k_{A}^{2}+C_{1}k_{z}^{2}+C_{2}k_\parallel^2 , \\
\mathbf{h}_s\left(\mathbf{k}\right) =&\left(sv_sk_x,\;-v_sk_y,\;M_0-M_{2}k_{A}^{2}-M_1k_z^2-M_2k_\parallel^2-s\lambda\right),
\end{split}
\end{eqnarray}
where $v_s=v_0-sv_A$. In the above equations, the terms related to $k_A$ describe the effects of the off-resonant light. Specifically, there are two main effects induced by the off-resonant light. One is that the parameters $C_0$ and $M_0$ are modified by the new terms $C_{2}k_{A}^{2}$ and $-M_{2}k_{A}^{2}$ respectively. Noting that these terms are spin-independent and the energy bands of the system are still spin-degenerate. Another effect is that new spin-dependent terms $-s\lambda \tau_{z}-v_{A}\left( k_{x}\tau _{x}-sk_{y}\tau _{y}\right)$ are generated to destroy the spin degeneracy. As a result, spin-dependent velocities $v_s=v_0-sv_A$ arise and the positions of Weyl points of different spin exhibit a different response to the light-field parameter $k_A$.
\par
Here, the driving frequency $\hbar\Omega $ and the bandwidth $BW$ are set as $\hbar\Omega=2{\rm eV} $ and $BW=0.24{\rm eV}$. The off-resonant condition is satisfied since $\hbar\Omega \gg BW$. The setting of the bandwidth, as well as the frequency, is reasonable since we only concern the low-energy behavior with the energy in the range of $|E|<0.12{\rm eV}$.

\section{ Derivation of the analytical RKKY interaction at the phase boundary between the WSM and the WHM}
Here, we drop the term $\epsilon\left(\mathbf{k}\right)$ for facilitating the calculation of the analytical results. For the phase boundary between the WSM and the WHM, i.e., at the critical point $k_A=k_{c}=\sqrt{M_{0}/\left[M_{2}-v_{0}^{2}/\left(\hbar\Omega\right)\right]}$, the spin-up band is in the WSM phase while the spin-down one is in the S-DSM phase. For the spin-up band, the bulk conduction and valence bands touch each other at two Weyl points located at $(0, 0, \pm k_{0,+})$ with $k_{0,+}=\sqrt{\left(M_{0}-M_{2}k_{A}^{2}-\lambda \right)/ M_{1}}$. In this case, one can linearize the Hamiltonian $H_+$ of the Eq. (\ref{I7}) around the Weyl points $(0, 0, \pm k_{0,+})$ to the following low-energy model
\begin{eqnarray}\label{II1}
\begin{split}
 H_{+,\eta}=\left(
       \begin{array}{cc}
        \eta v_{z}k_{z}^{\prime } & v_{+}\left( k_{x}+ik_{y}\right) \\
v_{+}\left( k_{x}-ik_{y}\right) & -\eta v_{z}k_{z}^{\prime }
       \end{array}
     \right),
 \end{split}
\end{eqnarray}
where $v_{z}=-2M_{1}k_{0,+}$ and $\eta=\pm$ denote the chirality of the two Weyl points.  For the spin-down band, the Hamiltonian $H_-$ in the Eq. (\ref{I7}) can be further simplified as
\begin{eqnarray}\label{II2}
\begin{split}
 H_{-}=\begin{pmatrix}
-M_{1}k_{z}^{2} & -v_{-}\left( k_{x}-ik_{y}\right)  \\
-v_{-}\left( k_{x}+ik_{y}\right)  & M_{1}k_{z}^{2}%
\end{pmatrix}.
 \end{split}
\end{eqnarray}
\par
\subsection{The case with impurities deposited on the line connecting the Weyl points}
For impurities in the $z$ axis [i.e., $\mathbf{R}=(0,0,R_z)$], according to the Eqs. (\ref{m5}) and (\ref{II1}), the Green's function of the spin-up band can be calculated as
\begin{eqnarray}\label{I3}
\begin{split}
G_{+}\left( \mathbf{R},\omega \right) =&\sum\limits_{\eta =\pm }\frac{1}{\left( 2\pi \right) ^{3}}\int \int \int
dk_{x}dk_{y}dk^\prime_{z}e^{i\left( k^\prime_{z}+\eta k_{0,+}\right) R_{z}}\frac{1}{\omega ^{2}-v_{+}^{2}k_{\parallel }^{2}-v_{z}^{2}{k^\prime_{z}}^{ 2}}\begin{pmatrix}
\omega +\eta v_{z}k^\prime_{z} & v_{+}\left( k_{x}+ik_{y}\right) \\
v_{+}\left( k_{x}-ik_{y}\right) & \omega -\eta v_{z}k^\prime_{z}\end{pmatrix},\\
=&\sum\limits_{\eta =\pm }e^{i\eta  k_{0,+}R_{z}}\frac{1}{\left(
2\pi \right) ^{3}}\int k_{\parallel }dk_{\parallel }\int dk^\prime_{z}e^{ik^\prime_{z}R_{z}}\int_{0}^{2\pi }d\theta _{\parallel }\frac{1}{\omega ^{2}-v_{+}^{2}k_{\parallel }^{2}-v_{z}^{2}{k^\prime_{z}}^{ 2}}\begin{pmatrix}
\omega +\eta v_{z}k^\prime_{z} & v_{+}k_{\parallel }e^{i\theta
_{\parallel }} \\
v_{+}k_{\parallel }e^{-i\theta _{\parallel }} & \omega -\eta v_{z}k^\prime_{z}%
\end{pmatrix}, \\
=&\sum\limits_{\eta =\pm }e^{i\eta k_{0,+}R_{z}}\frac{2}{\left(
2\pi \right) ^{2}}\int_{0}^{\infty }dk_{\parallel }\int_{0}^{\infty
}dk^\prime_{z}k_{\parallel }\frac{1}{\omega ^{2}-v_{+}^{2}k_{\parallel }^{2}-v_{z}^{2}{k^\prime_{z}}^{ 2}}\begin{pmatrix}
\omega \cos \left(k^\prime_{z}R_{z}\right) +i\eta v_{z}k^\prime_{z}\sin \left( k^\prime_{z}R_{z}\right) & 0 \\
0 & \omega \cos \left( k^\prime_{z}R_{z}\right) -i\eta v_{z}k^\prime_{z}\left( k^\prime_{z}R_{z}\right)%
\end{pmatrix},\\
\end{split}
\end{eqnarray}
Applying a parameter transformation $v_+k_{\parallel}=v_{z}k_{\parallel}^{\prime}$, the above Green's function can be further simplified as
\begin{eqnarray}\label{II4}
\begin{split}
G_{+}\left( \mathbf{R},\omega \right)=&\sum\limits_{\eta =\pm }e^{i\eta k_{0,+}R_{z}}\frac{2}{\left(
2\pi \right) ^{2}}\frac{v_{z}^{2}}{v_{+}^{2}}\int_{0}^{\infty }dk_{\parallel
}^{\prime }\int_{0}^{\infty }dk^\prime_{z} k_{\parallel }^{\prime }\frac{%
\begin{pmatrix}
\omega \cos \left( k^\prime_{z}R_{z}\right) +i\eta v_{z}k_{z}^{\prime}\sin \left(k^\prime_{z}R_{z}\right) & 0 \\
0 & \omega \cos \left( k^\prime_{z}R_{z}\right) -i\eta v_{z}k^\prime_{z}\left( k^\prime_{z}R_{z}\right)%
\end{pmatrix}%
}{\omega ^{2}-v_{z}^{2}k_{\parallel }^{\prime 2}-v_{z}^{2}{k^\prime_{z}}^{2}},\\
=&\sum\limits_{\eta =\pm }e^{i\eta k_{0,+}R_{z}}\frac{2v_{z}^{2}}{%
\left( 2\pi \right) ^{2}v_{+}^{2}}\int_{0}^{\infty }k^{2}dk\frac{%
\begin{pmatrix}
\omega \frac{\sin \left( kR_{z}\right) }{kR_{z}}+i\eta v_{z}k\frac{\sin
\left( kR_{z}\right) -kR_{z}\cos \left( kR_{z}\right) }{k^{2}R_{z}^{2}} & 0
\\
0 & \omega \frac{\sin \left( kR_{z}\right) }{kR_{z}}-i\eta v_{z}k\frac{\sin
\left( kR_{z}\right) -kR_{z}\cos \left( kR_{z}\right) }{k^{2}R_{z}^{2}}%
\end{pmatrix}}{\omega ^{2}-v_{z}^{2}k^{2}}, \\
=&\begin{pmatrix}
r_{+}+t_{+} & 0 \\
0 & r_{+}-t_{+}%
\end{pmatrix},\\
\end{split}
\end{eqnarray}
with
\begin{eqnarray}\label{II5}
\begin{split}
r_{+}=&\frac{-\omega }{2\pi
v_{+}^{2}}\frac{\cos \left( k_{0,+}R_{z}\right) }{R_{z}}e^{i\frac{%
R_{z}\omega }{v_{z}}},\\
t_{+}=&-\frac{v_{z}\sin
\left( k_{0,+}R_{z}\right) }{2\pi v_{+}^{2}R_{z}^{2}}\left( \frac{%
i\omega R_{z}}{v_{z}}-1\right) e^{i\frac{R_{z}\omega }{v_{z}}}.
\end{split}
\end{eqnarray}
Similarly, one can calculate $G_{+}\left( -\mathbf{R},\omega \right)$, which satisfies $G_{+}\left( -\mathbf{R},\omega \right)=G_{+}\left( \mathbf{R},\omega \right)$.
\par
According to the Eqs. (\ref{m5}) and (\ref{II2}), the Green's function of the spin-down band can be calculated as
\begin{eqnarray}\label{II6}
\begin{split}
G_{-}\left( \mathbf{R},\omega \right) =&\frac{1}{\left( 2\pi \right) ^{3}}%
\int \int \int dk_{x}dk_{y}dk_{z}e^{ik_{z}R_{z}}\frac{%
\begin{pmatrix}
\omega -M_{1}k_{z}^{2} & -v_{-}\left( k_{x}-ik_{y}\right) \\
-v_{-}\left( k_{x}+ik_{y}\right) & \omega +M_{1}k_{z}^{2}%
\end{pmatrix}%
}{\omega ^{2}-v_{-}^{2}k_{\parallel }^{2}-M_{1}^{2}k_{z}^{4}}, \\
=&\frac{1}{\left( 2\pi \right) ^{3}}\int_{0}^{\infty }k_{\parallel
}dk_{\parallel }\int dk_{z}e^{ik_{z}R_{z}}\int_{0}^{2\pi }d\theta
_{\parallel }\frac{%
\begin{pmatrix}
\omega -M_{1}k_{z}^{2} & -v_{-}k_{\parallel }e^{-i\theta _{\parallel }} \\
-v_{-}k_{\parallel }e^{i\theta _{\parallel }} & \omega +M_{1}k_{z}^{2}%
\end{pmatrix}%
}{\omega ^{2}-v_{-}^{2}k_{\parallel }^{2}-M_{1}^{2}k_{z}^{4}}, \\
=&\frac{2}{\left( 2\pi \right) ^{2}}\int_{0}^{\infty }k_{\parallel
}dk_{\parallel }\int_{0}^{\infty } dk_{z}\cos\left(k_{z}R_{z}\right)\frac{%
\begin{pmatrix}
\omega -M_{1}k_{z}^{2} & 0 \\
0 & \omega +M_{1}k_{z}^{2}%
\end{pmatrix}%
}{\omega ^{2}-v_{-}^{2}k_{\parallel }^{2}-M_{1}^{2}k_{z}^{4}}, \\
=&\frac{1}{4\pi ^{2}}\int_{0}^{\infty }dk_{\parallel }\int_{0}^{\infty
}dk_{z}^{\prime }\frac{k_{\parallel }\cos \left( \sqrt{k_{z}^{\prime }}%
R_{z}\right) }{\sqrt{k_{z}^{\prime }}}\frac{%
\begin{pmatrix}
\omega -M_{1}k_{z}^{\prime } & 0 \\
0 & \omega +M_{1}k_{z}^{\prime }%
\end{pmatrix}%
}{\omega ^{2}-v_{-}^{2}k_{\parallel }^{2}-M_{1}^{2}k_{z}^{\prime 2}}, \\
=&\frac{M_{1}^{2}}{4\pi ^{2}v_{-}^{2}}\int_{0}^{\infty }dk_{\parallel
}^{\prime }\int_{0}^{\infty }dk_{z}^{\prime }\frac{k_{\parallel }^{\prime
}\cos \left( \sqrt{k_{z}^{\prime }}R_{z}\right) }{\sqrt{k_{z}^{\prime }}}%
\frac{%
\begin{pmatrix}
\omega -M_{1}k_{z}^{\prime } & 0 \\
0 & \omega +M_{1}k_{z}^{\prime }%
\end{pmatrix}%
}{\omega ^{2}-M_{1}^{2}k_{\parallel }^{\prime 2}-M_{1}^{2}k_{z}^{\prime 2}}, \\
=&\frac{M_{1}^{2}}{4\pi ^{2}v_{-}^{2}}\int_{0}^{\infty }\frac{k^{3/2}}{%
\omega ^{2}-M_{1}^{2}k^{2}}dk\int_{0}^{\frac{\pi }{2}}d\theta \frac{\cos
\left( \theta \right) \cos \left[ \sqrt{k\sin \left( \theta \right) }R_{z}%
\right] }{\sqrt{\sin \left( \theta \right) }}%
\begin{pmatrix}
\omega -M_{1}k\sin \left( \theta \right)  & 0 \\
0 & \omega +M_{1}k\sin \left( \theta \right)
\end{pmatrix},\\
=&\frac{M_{1}^{2}\omega }{2\pi ^{2}v_{-}^{2}R_{z}}\int_{0}^{\infty }\frac{%
k\sin \left( \sqrt{k}R_{z}\right) }{\omega ^{2}-M_{1}^{2}k^{2}}dk%
\begin{pmatrix}
1 & 0 \\
0 & 1%
\end{pmatrix}%
+\frac{M_{1}^{3}}{2\pi ^{2}v_{-}^{2}R_{z}^{3}}\int_{0}^{\infty }\frac{k\sin
\left( \sqrt{k}R_{z}\right) \left( 2-kR_{z}^{2}\right) -2R_{z}k^{3/2}\cos
\left( \sqrt{k}R_{z}\right) }{\omega ^{2}-M_{1}^{2}k^{2}}dk%
\begin{pmatrix}
1 & 0 \\
0 & -1%
\end{pmatrix}, \\
=&\begin{pmatrix}
r_{-}+t_{-} & 0 \\
0 & r_{-}-t_{-}%
\end{pmatrix},
\end{split}
\end{eqnarray}
where
\begin{eqnarray}\label{II7}
\begin{split}
r_{-}=& -\frac{\omega }{4\pi R_{z}v_{-}^{2}}\left( e^{iR_{z}\sqrt{\frac{\omega }{-M_1}}}+e^{-R_{z}\sqrt{\frac{\omega }{-M_1}}}\right),\\
t_{-}=&e^{iR_{z}\sqrt{\frac{\omega }{-M_{1}}}}\frac{-2iR_{z}\sqrt{-M_{1}\omega }%
-R_{z}^{2}\omega -2M_{1}}{4\pi v_{-}^{2}R_{z}^{3}}+e^{-R_{z}\sqrt{\frac{%
\omega }{-M_{1}}}}\frac{2R_{z}\sqrt{-M_{1}\omega }+R_{z}^{2}\omega -2M_{1}}{%
4\pi v_{-}^{2}R_{z}^{3}}.
\end{split}
\end{eqnarray}
Similarly, one can calculate $G_{-}\left( -\mathbf{R},\omega \right)$, which satisfies $G_{-}\left( -\mathbf{R},\omega \right)=G_{-}\left( \mathbf{R},\omega \right)$.
\par
Plugging the Green's functions of Eqs. (\ref{II4}) and (\ref{II6}) into the Eq. (\ref{m4})  of the main text and summing over the spin and orbital degrees of freedom, the RKKY components can be written in the form of
\begin{eqnarray}\label{II8}
\begin{split}
H_{\rm R}=J_{xx}\left( S_{1}^{x}S_{2}^{x}+S_{1}^{y}S_{2}^{y}\right)
+J_{zz}S_{1}^{z}S_{2}^{z},
\end{split}
\end{eqnarray}
with
\begin{eqnarray}\label{II9}
\begin{split}
J_{xx}=&\frac{-4\lambda ^{2}}{\pi }{\rm{Im}}\int_{-\infty }^{0}\left(
r_{+}r_{-}+t_{+}t_{-}\right) d\omega, \\
J_{zz}=&\frac{-2\lambda ^{2}}{%
\pi }{\rm{Im}}\int_{-\infty }^{0}\left(
r_{+}^{2}+r_{-}^{2}+t_{+}^{2}+t_{-}^{2}\right) d\omega .
\end{split}
\end{eqnarray}
Plugging the Eqs. (\ref{II5}) and (\ref{II7}) into the above equation, $J_{xx}$ can be calculated as
\begin{eqnarray}\label{II10}
\begin{split}
J_{xx}=& -{\rm Im}\int_{-\infty }^{0}d\omega \frac{e^{\frac{iR_{z}\omega }{%
v_{z}}}\left( e^{iR_{z}\sqrt{\frac{\omega }{-M_{1}}}}+e^{-R_{z}\sqrt{\frac{%
\omega }{-M_{1}}}}\right) \omega ^{2}}{2\pi
^{3}R_{z}^{2}v_{-}^{2}v_{+}^{2}/\left[ \lambda ^{2}\cos \left(
k_{0,+}R_{z}\right) \right] }-{\rm Im}\int_{-\infty }^{0}d\omega \frac{%
e^{R_{z} \left( \frac{i\omega}{v_{z}}-\sqrt{\frac{\omega}{-M_{1} }}\right)
}\left( v_{z}-iR_{z}\omega \right) \left[ R_{z}\left( R_{z}\omega +2\sqrt{%
-M_{1}\omega }\right) -2M_{1}\right] }{2\pi ^{3}R_{z}^{5}v_{-}^{2}v_{+}^{2}/%
\left[ \lambda ^{2}\sin \left( k_{0}R_{z}\right) \right] }\\
&+{\rm Im}\int_{-\infty }^{0}d\omega \frac{e^{iR_{z} \left( \frac{\omega}{%
v_{z}}+\sqrt{\frac{\omega}{-M_{1}}}\right) }\left( v_{z}-iR_{z}\omega
\right) \left[ R_{z}\left( R_{z}\omega+i2\sqrt{-M_{1}\omega }\right) +2M_{1}%
\right] }{2\pi ^{3}R_{z}^{5}v_{-}^{2}v_{+}^{2}/\left[ \lambda ^{2}\sin
\left( k_{0}R_{z}\right) \right] },\\
=&\frac{\left( R_{z}^{3}v_{z}^{3}-32M_{1}^{2}R_{z}v_{z}\right) \cot \left(
k_{0}R_{z}\right) +2M_{1}\left( 9v_{z}^{2}R_{z}^{2}-32M_{1}^{2}\right) }{%
16\pi ^{3}M_{1}^{2}R_{z}^{6}v_{-}^{2}v_{+}^{2}/\left[ v_{z}^{2}\sin \left(
k_{0}R_{z}\right) \right] }+{\rm Im}\left\{ \frac{\left( 1+i\right) e^{-%
\frac{iR_{z}v_{z}}{4M_{1}}}\left(
v_{z}^{2}R_{z}^{2}-60M_{1}^{2}+i20M_{1}R_{z}v_{z}\right) {\rm erfc}\left[
\frac{\left( -1\right) ^{1/4}\sqrt{R_{z}v_{z}}}{2\sqrt{-M_{1}}}\right] }{64%
\sqrt{2}M_{1}^{5/2}\pi ^{5/2}R_{z}^{9/2}v_{-}^{2}v_{+}^{2}/\left[
v_{z}^{7/2}\cos \left( k_{0}R_{z}\right) \right] }\right\} \\
&+{\rm Im}\left\{ \frac{\left( 1-i\right) e^{-\frac{iR_{z}v_{z}}{4M_{1}}%
}\left( v_{z}^{2}R_{z}^{2}-28M_{1}^{2}+i20M_{1}R_{z}v_{z}\right) {\rm erfc}%
\left[ \frac{\left( -1\right) ^{1/4}\sqrt{R_{z}v_{z}}}{2\sqrt{-M_{1}}}\right]
+\left( 1+i\right) e^{\frac{iR_{z}v_{z}}{4M_{1}}}\left(
v_{z}^{2}R_{z}^{2}-28M_{1}^{2}-i20M_{1}R_{z}v_{z}\right) {\rm erfc}\left[
\frac{\left( -1\right) ^{3/4}\sqrt{R_{z}v_{z}}}{-2\sqrt{-M_{1}}}\right] }{64%
\sqrt{2}M_{1}^{5/2}\pi ^{5/2}R_{z}^{9/2}v_{-}^{2}v_{+}^{2}/\left[
v_{z}^{7/2}\sin \left( k_{0}R_{z}\right) \right] }\right\} \\
&-{\rm Im}\left\{ \frac{\left( 1-i\right) e^{\frac{iR_{z}v_{z}}{4M_{1}}%
}\left( 60M_{1}^{2}+i20M_{1}R_{z}v_{z}-v_{z}^{2}R_{z}^{2}\right) {\rm erfc}%
\left[ \frac{\left( -1\right) ^{3/4}\sqrt{R_{z}v_{z}}}{-2\sqrt{-M_{1}}}%
\right] }{64\sqrt{2}M_{1}^{5/2}\pi ^{5/2}R_{z}^{9/2}v_{-}^{2}v_{+}^{2}/\left[
v_{z}^{7/2}\cos \left( k_{0}R_{z}\right) \right] }\right\},
\end{split}
\end{eqnarray}
where ${\rm erfc}(t)$ is the complementary error function. Similarly, $J_{zz}$ can be calculated as
\begin{eqnarray}\label{II11}
\begin{split}
J_{zz}=&{\rm Im}\int_{-\infty }^{0}e^{\frac{i2R_{z}\omega }{v_{z}}}\frac{\left[ v_{z}-\left( 1+i\right) R_{z}\omega \right] \left[ v_{z}+\left(1-i\right) R_{z}\omega \right] \cos \left( 2k_{0,+}R_{z}\right) -v_{z}\left(v_{z}-i2R_{z}\omega \right) }{4\pi ^{3}R_{z}^{4}v_{+}^{4}}d\omega -{\rm Im}\int_{-\infty }^{0}\frac{\left( e^{iR_{z}\sqrt{\frac{\omega }{-M_1}}}+e^{-R_{z}\sqrt{\frac{\omega }{-M_1}}}\right)
^{2}\omega ^{2}}{8\pi ^{3}R_{z}^{2}v_{-}^{4}}%
d\omega\\
&-{\rm Im}\int_{-\infty }^{0}\frac{\left[ e^{-R_{z}\sqrt{\frac{\omega }{-M_{1}}}}\left(2R_{z}\sqrt{-M_{1}\omega }+R_{z}^{2}\omega -2M_{1}\right)-e^{iR_{z}\sqrt{\frac{\omega }{-M_{1}}}}\left(2iR_{z}\sqrt{-M_{1}\omega }%
+R_{z}^{2}\omega +2M_{1}\right) \right] ^{2}}{8\pi ^{3}R_{z}^{6}v_{-}^{4}}%
d\omega ,\\
=&\frac{v_{z}^{3}\left[ 2-3\cos \left( 2k_{0,+}R_{z}\right) \right] }{8\pi
^{3}R_{z}^{5}v_{+}^{4}}-\frac{12M_{1}^{3}}{\pi ^{3}R_{z}^{8}v_{-}^{4}}.
\end{split}
\end{eqnarray}
\par
\subsection{The case with impurities deposited on the plane perpendicular to the line connecting the Weyl points}
For impurities deposited on the $x$-$y$ plane [i.e., $\mathbf{R}=(R_x,R_y,0)$], according to the Eqs. (\ref{m5}) and (\ref{II1}), the Green's function of the spin-up band can be calculated as
\begin{eqnarray}\label{II12}
\begin{split}
G_{+}\left( \mathbf{R},\omega \right) =&\sum\limits_{\eta =\pm }\frac{1}{\left( 2\pi \right) ^{3}}\int \int \int
dk_{x}dk_{y}dk^\prime_{z}e^{i\left( k_{x}R_{x}+k_{y}R_{y}\right) }\frac{1
}{\omega ^{2}-v_{+}^{2}k_{\parallel }^{2}-v_{z}^{2}{k^\prime_{z}}^{2}}\begin{pmatrix}
\omega +\eta v_{z}k^\prime_{z} & v_{+}\left( k_{x}+ik_{y}\right)  \\
v_{+}\left( k_{x}-ik_{y}\right)  & \omega -\eta v_{z}k^\prime_{z}\end{pmatrix}, \\
=&\frac{1}{2\pi ^{3}}\int_{0}^{\infty }k_{\parallel }dk_{\parallel
}\int_{0}^{\infty }dk^\prime_{z}\int_{0}^{2\pi }d\varphi e^{ik_{\parallel
}R_{\parallel }\cos \left( \varphi -\varphi _{R}\right) }\frac{1}{%
\omega ^{2}-v_{+}^{2}k_{\parallel }^{2}-v_{z}^{2}{k^\prime_{z}}^{2}}%
\begin{pmatrix}
\omega  & v_{+}k_{\parallel }e^{i\varphi } \\
v_{+}k_{\parallel }e^{-i\varphi } & \omega\end{pmatrix}, \\
=&\frac{1}{\pi ^{2}}\int_{0}^{\infty }k_{\parallel }dk_{\parallel
}\int_{0}^{\infty }dk^\prime_{z}\frac{1}{\omega ^{2}-v_{+}^{2}k_{\parallel
}^{2}-v_{z}^{2}{k^\prime_{z}}^{2}}%
\begin{pmatrix}
\omega J_{0}\left( k_{\parallel }R_{\parallel }\right)  & v_{+}ie^{i\varphi _{R}}k_{\parallel }J_{1}\left( k_{\parallel }R_{\parallel }\right)
\\
v_{+}ie^{-i\varphi _{R}}k_{\parallel }J_{1}\left( k_{\parallel
}R_{\parallel }\right)  & \omega J_{0}\left( k_{\parallel }R_{\parallel
}\right)
\end{pmatrix},
\end{split}
\end{eqnarray}
where $\varphi _{R}= \arctan(R_y/R_x)$. Applying a parameter transformation $v_{z}k^\prime_{z}=v_{+}k_{z}$, the above Green's function can be further simplified as
\begin{eqnarray}\label{II13}
\begin{split}
G_{+}\left( \mathbf{R},\omega \right) =&\frac{v_{+}}{\pi ^{2}v_{z}}\int_{0}^{\infty }k_{\parallel
}dk_{\parallel }\int_{0}^{\infty }dk_{z}\frac{1}{\omega
^{2}-v_{+}^{2}k_{\parallel }^{2}-v_{+}^{2}k_{z}^{ 2}}%
\begin{pmatrix}
\omega J_{0}\left( k_{\parallel }R_{\parallel }\right)  & v_{+}ie^{i\varphi _{R}}k_{\parallel }J_{1}\left( k_{\parallel }R_{\parallel }\right)
\\
v_{+}ie^{-i\varphi _{R}}k_{\parallel }J_{1}\left( k_{\parallel
}R_{\parallel }\right)  & \omega J_{0}\left( k_{\parallel }R_{\parallel
}\right)
\end{pmatrix},\\
=&\frac{v_{+}}{\pi ^{2}v_{z}}\int_{0}^{\infty }\frac{k^{2}}{\omega
^{2}-v_{+}^{2}k^{2}}dk\int_{0}^{\frac{\pi }{2}}d\theta \cos \left( \theta
\right)
\begin{pmatrix}
\omega J_{0}\left[ kR_{\parallel }\cos \left( \theta \right) \right]  &
v_{+}ie^{i\varphi _{R}}k\cos \left(\theta \right) J_{1}\left[
kR_{\parallel }\cos \left(\theta \right) \right]  \\
v_{+}ie^{-i\varphi _{R}}k\cos \left( \theta \right) J_{1}\left[
kR_{\parallel }\cos \left(\theta \right) \right]  & \omega J_{0}\left[
kR_{\parallel }\cos \left( \theta \right) \right]
\end{pmatrix}, \\
=& \frac{v_{+}}{\pi ^{2}v_{z}R_{\parallel }}\int_{0}^{\infty }\frac{k}{\omega
^{2}-v_{+}^{2}k^{2}}dk%
\begin{pmatrix}
\omega \sin \left( kR_{\parallel }\right)  & v_{+}ie^{i\varphi _{R}}
\left[ \sin \left( kR_{\parallel }\right) /R_{\parallel }-k\cos \left(
kR_{\parallel }\right) \right]  \\
v_{+}ie^{-i\varphi _{R}}\left[ \sin \left( kR_{\parallel }\right)
/R_{\parallel }-k\cos \left( kR_{\parallel }\right) \right]  & \omega \sin
\left( kR_{\parallel }\right)
\end{pmatrix}, \\
=&\begin{pmatrix}
r_{+} & e^{i\varphi _{\parallel}}q_{+} \\
e^{-i\varphi _{\parallel}}q_{+} & r_{+}%
\end{pmatrix},
\end{split}
\end{eqnarray}
where
\begin{eqnarray}\label{II14}
\begin{split}
r_{+}=-\frac{\omega }{2\pi v_{+}v_{z}R_{\parallel }}e^{i\frac{%
\omega R_{\parallel }}{v_{+}}},  \;\;\;\;\;\; q_{+}=-\frac{iv_{+}+\omega
R_{\parallel }}{2\pi v_{+}v_{z}R_{\parallel }^{2}}e^{i\frac{\omega
R_{\parallel }}{v_{+}}}.
\end{split}
\end{eqnarray}
Similarly, one can calculate $G_{+}\left( -\mathbf{R},\omega \right)$, which satisfies $G_{+}\left( -\mathbf{R},\omega \right)=G_{+}\left( \mathbf{R},\omega \right)|_{q_+\rightarrow -q_+}$.
\par
According to the Eqs. (\ref{m5}) and (\ref{II2}), the Green's function of the spin-down band can be calculated as
\begin{eqnarray}\label{II15}
\begin{split}
G_{-}\left( \mathbf{R},\omega \right) =&\frac{1}{\left( 2\pi \right) ^{3}}%
\int \int dk_{x}dk_{y}\int dk_{z}e^{i\left( k_{x}R_{x}+k_{y}R_{y}\right) }%
\frac{%
\begin{pmatrix}
\omega -M_{1}k_{z}^{2} & -v_{-}\left( k_{x}-ik_{y}\right) \\
-v_{-}\left( k_{x}+ik_{y}\right) & \omega +M_{1}k_{z}^{2}%
\end{pmatrix}%
}{\omega ^{2}-v_{-}^{2}k_{\parallel }^{2}-M_{1}^{2}k_{z}^{4}} ,\\
=&\frac{1}{\left( 2\pi \right) ^{3}}\int_{0}^{\infty }k_{\parallel
}dk_{\parallel }\int dk_{z}\int_{0}^{2\pi }d\varphi  e^{ik_{\parallel
}R_{\parallel }\cos \left( \varphi -\varphi _{R}\right) }\frac{%
\begin{pmatrix}
\omega -M_{1}k_{z}^{2} & -v_{-}k_{\parallel }e^{-i\varphi  } \\
-v_{-}k_{\parallel }e^{i\varphi } & \omega +M_{1}k_{z}^{2}%
\end{pmatrix}}{\omega ^{2}-v_{-}^{2}k_{\parallel }^{2}-M_{1}^{2}k_{z}^{4}}, \\
=&\frac{2}{\left( 2\pi \right) ^{2}}\int_{0}^{\infty }k_{\parallel
}dk_{\parallel }\int_{0}^{\infty }dk_{z}\frac{\begin{pmatrix}
\left( \omega -M_{1}k_{z}^{2}\right) J_{0}\left( k_{\parallel }R_{\parallel
}\right) & -v_{-}ie^{-i\varphi _{R}}k_{\parallel }J_{1}\left( k_{\parallel
}R_{\parallel }\right) \\
-v_{-}ie^{i\varphi _{R}}k_{\parallel }J_{1}\left( k_{\parallel }R_{\parallel
}\right) & \left( \omega +M_{1}k_{z}^{2}\right) J_{0}\left( k_{\parallel
}R_{\parallel }\right)%
\end{pmatrix}}{\omega ^{2}-v_{-}^{2}k_{\parallel }^{2}-M_{1}^{2}k_{z}^{4}}.
\end{split}
\end{eqnarray}
Applying a parameter transformation $-M_{1}k_{z}^{2}=v_{-}k_{z}^{\prime }$, the above Green's function can be further simplified as
\begin{eqnarray}\label{II16}
\begin{split}
G_{-}\left( \mathbf{R},\omega \right) =&\frac{\sqrt{v_{-}}}{\left( 2\pi \right) ^{2}\sqrt{-M_{1}}}\int_{0}^{\infty
}k_{\parallel }dk_{\parallel }\int_{0}^{\infty }dk_{z}^{\prime }\frac{1}{%
\sqrt{k_{z}^{\prime }}}\frac{%
\begin{pmatrix}
\left( \omega +v_{-}k_{z}^{\prime }\right) J_{0}\left( k_{\parallel
}R_{\parallel }\right)  & -v_{-}ie^{-i\varphi _{R}}k_{\parallel }J_{1}\left(
k_{\parallel }R_{\parallel }\right)  \\
-v_{-}ie^{i\varphi _{R}}k_{\parallel }J_{1}\left( k_{\parallel }R_{\parallel
}\right)  & \left( \omega -v_{-}k_{z}^{\prime }\right) J_{0}\left(
k_{\parallel }R_{\parallel }\right)
\end{pmatrix}%
}{\omega ^{2}-v_{-}^{2}k_{\parallel }^{2}-v_{-}^{2}k_{z}^{\prime 2}}, \\
=&\frac{\sqrt{v_{-}}}{\left( 2\pi \right) ^{2}\sqrt{-M_{1}}}%
\int_{0}^{\infty }k^{3/2}dk\int_{0}^{\frac{\pi }{2}}d\theta \frac{\cos
(\theta) }{\sqrt{\sin (\theta) }}\frac{%
\begin{pmatrix}
\left[ \omega +v_{-}k\sin \left(\theta\right) \right] J_{0}\left[ k\cos \left(\theta\right)
R_{\parallel }\right] & -v_{-}ie^{-i\varphi _{R}}k\cos \left(\theta\right) J_{1}\left[
k\cos \left(\theta\right) R_{\parallel }\right] \\
-v_{-}ie^{i\varphi _{R}}k\cos \left(\theta\right) J_{1}\left[ k\cos \left(\theta\right) R_{\parallel
}\right] & \left[ \omega -v_{-}k\sin\left( \theta\right) \right] J_{0}\left[ k\cos \left(\theta\right)
R_{\parallel }\right]\end{pmatrix}}{\omega ^{2}-v_{-}^{2}k^{2}}, \\
=&\frac{\sqrt{v_{-}}}{2\pi ^{2}\sqrt{-M_{1}}}\int_{0}^{\infty }%
\frac{k^{3/2}}{\omega ^{2}-v_{-}^{2}k^{2}}\begin{pmatrix}
_{0}F_{1}\left( ;\frac{5}{4};-\frac{k^{2}R_{\parallel }^{2}}{4}\right)
\omega +\frac{_{0}F_{1}\left( ;\frac{7}{4};-\frac{k^{2}R_{\parallel }^{2}}{4}%
\right) v_{-}k}{3} & -2v_{-}ie^{-i\varphi _{R}}R_{\parallel }\frac{%
_{0}F_{1}\left( ;\frac{9}{4};-\frac{k^{2}R_{\parallel }^{2}}{4}\right) k^{2}%
}{5} \\
-2v_{-}ie^{i\varphi _{R}}R_{\parallel }\frac{_{0}F_{1}\left( ;\frac{9%
}{4};-\frac{k^{2}R_{\parallel }^{2}}{4}\right) k^{2}}{5} & _{0}F_{1}\left( ;%
\frac{5}{4};-\frac{k^{2}R_{\parallel }^{2}}{4}\right) \omega -\frac{%
_{0}F_{1}\left( ;\frac{7}{4};-\frac{k^{2}R_{\parallel }^{2}}{4}\right) v_{-}k%
}{3}\end{pmatrix}dk, \\
=&\begin{pmatrix}
r_{-}+t_{-} & -e^{-i\varphi _{R}}q_{-} \\
-e^{i\varphi _{R}}q_{-} & r_{-}-t_{-}%
\end{pmatrix},
\end{split}
\end{eqnarray}
In the above equation, $_{0}F_{1}\left( ;b;t\right)$ is the confluent hypergeometric function, and $r_{-}$, $t_{-}$, $q_{-}$ read as
\begin{eqnarray}\label{II17}
\begin{split}
r_{-}=&\frac{\left( -1\right) ^{7/8}\sqrt{-v_{-}/M_{1}}\omega
^{5/4}K_{1/4}\left( -iR_{\parallel }\omega /v_{-}\right) \Gamma \left(
5/4\right) }{2^{3/4}\pi ^{2}v_{-}^{9/4}R_{\parallel }^{1/4}}, \\
t_{-}=&\frac{\left( -1\right) ^{5/8}\omega ^{3/4}K_{3/4}\left(
-iR_{\parallel }\omega /v_{-}\right) \Gamma \left( 7/4\right) }{3\times
2^{1/4}\sqrt{-M_{1}}\pi ^{2}v_{-}^{5/4}R_{\parallel }^{3/4}}, \\
q_{-}=&-\frac{\omega ^{5/4}K_{5/4}\left( -iR_{\parallel }\omega /v_{-}\right)
\Gamma \left( 5/4\right) }{\left( 1-i\right) ^{3/2}M_{1}^{1/2}\pi
^{2}v_{-}^{7/4}R_{\parallel }^{1/4}},
\end{split}
\end{eqnarray}
where $K_{v}\left(t\right) $ is the modified Bessel function of the second kind and $\Gamma \left( t\right) $ is the Gamma function. Similarly, one can calculate $G_{-}\left( -\mathbf{R},\omega \right)$, which satisfies $G_{-}\left( -\mathbf{R},\omega \right)=G_{-}\left( \mathbf{R},\omega \right)|_{q_-\rightarrow -q_-}$.
\par
Plugging the Green's function of Eqs. (\ref{II13}) and (\ref{II16}) into the Eq. (\ref{m4}) of the main text and summing over the spin and orbital degrees of
freedom, the RKKY components can be written as $H_{\rm R}=J_{xx}\left( S_{1}^{x}S_{2}^{x}+S_{1}^{y}S_{2}^{y}\right)
+J_{zz}S_{1}^{z}S_{2}^{z}$, where $J_{xx}$ and $J_{zz}$ read as
\begin{eqnarray}\label{II18}
\begin{split}
J_{xx}=&\frac{-4\lambda ^{2}}{\pi }{\rm {Im}}\int_{-\infty }^{u_{F}}\left[
r_{+}r_{-}+q_{-}q_{+}\cos \left( 2\varphi _R\right) %
\right] d\omega, \\
J_{zz}=&\frac{-2\lambda ^{2}}{\pi }{\rm {Im}}\int_{-\infty }^{u_{F}}\left(
r_{+}^{2}-q_{+}^{2}+t_{-}^{2}+r_{-}^{2}-q_{-}^{2}\right) d\omega.
\end{split}
\end{eqnarray}
Plugging the Eqs. (\ref{II14}) and (\ref{II17}) to the RKKY component $J_{xx}$ of the above equation, $J_{xx}$ can be solved as
 \begin{eqnarray}\label{II19}
\begin{split}
J_{xx}=&-{\rm Im}\int_{-\infty }^{0}e^{\frac{iR_{\parallel }\omega }{v_{+}}}\frac{%
\lambda ^{2}\Gamma \left( 5/4\right) \omega ^{5/4}\sqrt{1-i}}{\pi
^{4}R_{\parallel }^{5/4}v_{-}^{7/4}v_{+}v_{z}\sqrt{-M_1}}\left[ \omega K_{1/4}\left( -iR_{\parallel }\omega /v_{-}\right)+\frac{\left( iv_{+}+R_{\parallel }\omega \right)
K_{5/4}\left( -iR_{\parallel }\omega /v_{-}\right)\cos\left(2\varphi_R\right) }{R_{\parallel }}\right] d\omega , \\
=&\frac{1}{R_{\parallel }^{4.5}}\frac{1024\sqrt{\pi}v_+^{7/2}\left(v_+^2-v_-^2\right)^{1/4}\left(5v_-^2+2v_+^2\right)
\Gamma\left(2.75\right)+1575\left(v_-^2-v_+^2\right)^3\Gamma\left(-3.75\right)\left[_{2}F_{1}\left(3,3.5;3.75;\frac{v_++v_-}{2v_+}\right)\right]}
{896\sqrt{2}\pi^4M_1v_+\left(v_+^2-v_-^2\right)^3v_z/\left[\lambda^2\sqrt{-M_1}v_-^{3/2}\Gamma\left(1.25\right)\right]} \\
&+\frac{1}{R_{\parallel }^{4.5}}\frac{1024\sqrt{\pi}v_+^{9/2}\left(v_+^2-v_-^2\right)^{1/4}\left(2v_+^2-9v_-^2\right)\Gamma\left(1.75\right)+225
\left(v_-^2-v_+^2\right)^3\Gamma\left(-3.75\right)}{512\sqrt{2}\pi^4\sqrt{-M_1}\left(v_-^2-v_+^2\right)^3v_+v_z
/\left[\sqrt{v_-}\Gamma\left(1.25\right)\right]}\cos\left(2\varphi_R\right) \\
&+\frac{1}{R_{\parallel }^{4.5}}\frac{11v_+\left[_{2}F_{1}\left(1,3.5;2.75;\frac{v_-+v_+}{2v_+}\right)\right]-7v_-\left[_{2}F_{1}\left(2,4.5;3.75;\frac{v_-+v_+}{2v_+}\right)\right]}{512\sqrt{2}\pi^4\sqrt{-M_1}\left(v_-^2-v_+^2\right)^3v_+v_z
/\left[\sqrt{v_-}\Gamma\left(1.25\right)\right]}\cos\left(2\varphi_R\right),
\end{split}
\end{eqnarray}
where $_{2}F_{1}\left(a,b;c;t\right)$ is the Gauss hypergeometric function. Similarly, $J_{zz}$ can be solved as
\begin{eqnarray}\label{II20}
\begin{split}
J_{zz}=&{\rm Im}\int_{-\infty }^{0}\left\{ \frac{K_{1/4}^{2}\left(
-iR_{\parallel }\omega /v_{-}\right) -K_{5/4}^{2}\left( -iR_{\parallel
}\omega /v_{-}\right) }{2M_{1}\pi ^{5}R_{\parallel }^{1/2}v_{+}^{7/2}/\left[
\lambda ^{2}\left( 1-i\right) \omega ^{5/2}\Gamma ^{2}\left( \frac{5}{4}\right) \right] }%
-\frac{\lambda ^{2}\left( 1+i\right) \omega ^{3/2}K_{3/4}^{2}\left( -iR_{\parallel
}\omega /v_{-}\right) \Gamma ^{2}\left( \frac{7}{4}\right) }{9M_{1}\pi
^{5}R_{\parallel }^{3/2}v_{+}^{5/2}}+\frac{\lambda ^{2}\left( i2R_{\parallel }\omega
-v_{+}\right) e^{i2R_{\parallel }\omega /v_{+}}}{2\pi ^{3}R_{\parallel
}^{4}v_{+}v_{z}^{2}}\right\} d\omega, \\
=&-\frac{\lambda ^{2}}{15M_{1}\pi ^{3}}\frac{1}{R_{\parallel }^{4}}+\frac{v_{+}\lambda ^{2}}{2\pi
^{3}v_{z}^{2}}\frac{1}{R_{\parallel }^{5}}.
\end{split}
\end{eqnarray}

\end{widetext}

\end{document}